\documentclass[a4paper,useAMS,fleqn,usenatbib]{mnras}

\usepackage[T1]{fontenc}
\usepackage{ae,aecompl}

\usepackage{dingbat}

\usepackage{graphicx}	
\usepackage{amsmath}	
\usepackage{amssymb}	
\usepackage{color}
\usepackage{colortbl}
\usepackage[authoryear]{natbib}
\usepackage[normalem]{ulem} 

\definecolor{LightGray}{gray}{0.9}
\definecolor{LightGray1}{gray}{0.8}
\definecolor{darkforestgreen}{RGB}{37,65,23}
\definecolor{ferngreen}{RGB}{102,124,38}

\def\degr{\hbox{$^\circ$}}

\def\farcs{\hbox{$.\!\!^{\prime\prime}$}}

\usepackage{pifont} 
\DeclareFontFamily{U}{astrosym}{}
\DeclareFontShape{U}{astrosym}{m}{n}{<-> astrosym}{}

\title[New orbit recalculations of comet C/1890 F1 Brooks and its dynamical evolution]{New orbit recalculations of comet C/1890~F1 Brooks and its dynamical evolution}
\author[M. Kr\'{o}likowska and P.A. Dybczy\'{n}ski ]{Ma\l gorzata Kr\'{o}likowska$^1$\thanks{E-mail:
mkr@cbk.waw.pl} and Piotr A. Dybczy\'{n}ski$^2$, \thanks{E-mail: dybol@amu.edu.pl} \\
$^1$Space Research Centre of the Polish Academy of Sciences (CBK PAN),
Bartycka 18A, 00-716 Warsaw, Poland \\
$^2$Astronomical Observatory Institute, Faculty of Physics,
A.~Mickiewicz University,
S\l oneczna 36, 60-286 Pozna\'{n}, Poland\\
}

\date{Accepted XXX. Received YYY; in original form ZZZ}

\pubyear{2015}

\begin{document}
\label{firstpage}
\pagerange{\pageref{firstpage}--\pageref{lastpage}}
\maketitle


\begin{abstract}
\noindent C/1890~F1 Brooks belongs to a group of nineteen comets used
by Jan Oort to support his famous hypothesis on the existence of
a spherical cloud containing hundreds of billions of comets with  
orbits of semimajor axes between 50 and 150\,thousand au.
Comet Brooks stands out from this group because of a long series of
astrometric observations as well as nearly two-year long observational
arc. Rich observational material makes this comet an ideal target for 
testing the rationality of an effort to recalculate astrometric
positions on the basis  of original (comet--star)-measurements using
modern star catalogues. This paper presents the results of such new
analysis based on two different  methods: ~(i)~automatic re-reduction
based on cometary positions and the (comet--star)-measurements, and
~(ii)~partially automatic re-reduction based  on the contemporary data
for originally used reference stars. We show that both methods offer
a significant reduction of orbital elements uncertainties. Based on
the most preferred orbital solution, the dynamical evolution of comet
Brooks  during three consecutive perihelion passages is discussed.
We conclude that C/1890~F1 is a  dynamically old comet that passed the
Sun at a distance below $5$\,au during its  previous perihelion passage.
Furthermore, its next perihelion passage will be a little closer than
during the 1890-1892 apparition. C/1890~F1 is interesting also because
it suffered extremely small planetary perturbations when it travelled
through the planetary zone. Therefore, in the next passage through
perihelion it will be once again a comet from the Oort spike. 
\end{abstract}



\section{Introduction}

\label{intro}

\noindent Jan Oort~\citeyearpar{oort:1950} based his historic hypothesis
on the existence of a distant, spherical cloud of cometary bodies
surrounding the Solar system on the original $1/a$-distribution of
nineteen near-parabolic comets with the best-determined (at that time)
original orbits. Comet C/1890~F1 Brooks belonged to this group.
It is distinguished by  a long and rich series of observations:
almost 2\,yr period covered by about 900~positional measurements.

A question of the distribution of the reciprocal original semimajor
axes $1/a_{\rm ori}$ is frequently repeated in the literature not
only to support the Oort Cloud (OC) hypothesis, but also to examine
the density distribution of the OC, recognized as a reservoir of
the cometary bodies that can potentially penetrate into the planetary
system \citep{fernandez_book:2005}. However, it was already
established \citep{yabushita:1989,dyb-hist:2001,kroli-dyb:2010,
dyb-kroli:2011}, that a significant fraction of investigated comets
with the original $1/a$ within the so-called Oort spike,
$1/a_{\rm ori}\le10^{-4}$\,au$^{-1}$, passed through the inner part
of the Solar system in their previous perihelion passage. Therefore,
the interpretation of the original $1/a$-distribution
of the near-parabolic comets should be treated with caution. In particular,
one should take into consideration that a vast majority of comets from
the Oort spike suffered planetary perturbations that have changed
significantly their semimajor axes during each deep passage through planetary zone.
Here we focus our investigation on the dynamical history of Oort spike
comets in the period extending back to their previous perihelion
passage. It is the only direct method that allows us to separate the
dynamically new comets of the Oort spike from the dynamically old ones.

Typical planetary perturbations defined by a change of $1/a$ during
the passage within the inner part of the Solar system are typically
two to four times greater than the present estimate of the Oort spike
width. In consequence, after visiting the inner planetary
system, a comet has an excessive chance to be outside the Oort spike, leaving
the Solar system on a hyperbolic orbit, or moving on a significantly
tighter orbit than previously. Since many comets discovered so
far will escape from the Solar system in the future, we also investigate
their future evolution which gives a much broader perspective on the
dynamical evolution of comets discovered as Oort spike comets 
\citep[hereafter Paper 5 and references therein]{dyb-kroli:2015}.

In addition to the above general arguments, there are also some specific
objectives of the present investigation. We chose comet C/1890~F1~Brooks
not only for its historic importance but also to develop and test
our methods of data treatment for such a long-ago discovered comet. In
the case of C/1890~F1, we have an opportunity to repeat the reduction
of astrometric observations using contemporary star catalogues, typically
much more precise, particularly in terms of proper motions. The 
(\begin{tiny}\Pisymbol{astrosym}{34}\end{tiny}\textendash \,$\star$)-measurements\footnote{(comet$-$star)-measurements}
($\Delta\alpha$ in right ascension and $\Delta\delta$
in declination) are fortunately published in original papers together with calculated
apparent comet positions  for almost all observations
of this comet. Apart from seven transit circle observations made in
Washington the only sources without explicit (\begin{tiny}\Pisymbol{astrosym}{34}\end{tiny}\textendash \,$\star$)-differences
are original papers reporting 99~observations from Bordeaux Observatory
(only one paper with five observations from Bordeaux listed also
(\begin{tiny}\Pisymbol{astrosym}{34}\end{tiny}\textendash \,$\star$)-measurements). Fortunately, these 
publications also consist of both, reference star identifications and their apparent places used for the reduction.
Thereby, the reconstruction of the (\begin{tiny}\Pisymbol{astrosym}{34}\end{tiny}\textendash \,$\star$)-measurements
is quite straightforward.

For all these reasons, we also decided to show in this paper how the determination of the 
observed osculating orbit%
\footnote{that is always the starting nominal orbit for our numerical calculations
of dynamical evolution}%
depends on the adopted method of data processing (how deep we are looking for sources of errors 
in the published data) and our choice of the catalogue for reference star positions. 
Two methods and two catalogues have been used for this purpose: PPM star
catalogue \citep{ppm:1988} and Tycho-2 star catalogue \citep{tycho2-cat:2000}.
 
The first, fully automatic, method is very useful in cases in which we have at hand a list of astrometric cometary positions 
together with (\begin{tiny}\Pisymbol{astrosym}{34}\end{tiny}\textendash \,$\star$)-measurements in right-ascension and declination, and we had not entered data\footnote{Large sets of such data typed in computer files were prepared during a
long campaign of collecting observations of one-apparition comets as a part of the Warsaw Catalogue of Cometary Orbits project \citep{krol-sit-et-al:2014}.} on the reference stars originally used by observers. 
It involves the reconstruction of hypothetical star positions on the basis of (\begin{tiny}\Pisymbol{astrosym}{34}\end{tiny}\textendash \,$\star$)-measurements, and published cometary positions if they were given by observers. Otherwise, the cometary positions are estimated from preliminary orbital elements. Next, the re-reduction of these stars positions using PPM catalogue is carried out. Finally, the positions of comets are calculated based on obtained present-day star positions and
original (\begin{tiny}\Pisymbol{astrosym}{34}\end{tiny}\textendash \,$\star$)-measurements.  
Of course, it can also be applied to a star catalogue other than PPM, but here the PPM catalogue is used for a comparison purpose. This method was successfully used for many comets discovered long ago (\citealp{gabryszewski:1997} and \citealp{krol-sit-et-al:2014}), thus it is only very briefly described in section~\ref{stars_in_PPM}.

In this investigation however, all the data originally published by observers were collected. Therefore, a much more in-depth analysis of the original data can be done and the second method described in this paper is devoted
to such an investigation. The Tycho-2 catalogue is chosen as it is perceived to be one of the best modern catalogues for this purpose, particularly regarding proper motions. In the first step, we collected a list of mean coordinates of all reference stars used by all observers and then their contemporary astrometric data were automatically obtained. Next, we calculated astrometric positions of the comet using 
(\begin{tiny}\Pisymbol{astrosym}{34}\end{tiny}\textendash \,$\star$)-measurements.
Finally, manual and time consuming part was to investigate the causes of large residuals obtained from this automated
step for about 30~per cent of all observations. About half of them resulted from the wrong star identifications and the rest came from some inconsistency or typographical errors in papers published by observers.
More details about this second method are given in section~\ref{section:stars_in_Tycho-2}.

The paper consists of seven sections. Next section describes
the observational material and previous orbital determinations
of C/1890~F1~Brooks that exist in literature. In section~\ref{section:star_recalculation}
details of the method of reference star position recalculation
using the procedures based on PPM and Tycho-2 star catalogues are given, and
a brief description of further steps of our data treatment is delivered.

\noindent A grid of osculating orbits based on both methods of reference
star recalculations is described in Section~\ref{section:our_osculating_orbits},
where also the analysis of differences between these results is presented.
The original and future orbits are discussed in Section~\ref{section:original_future_orbits}.
The final result of this study, the most important from the dynamical
point of view, is presented in Section~\ref{section:past_next_orbits},
where we discuss what can be deduced about the origin of comet C/1890~F1~Brooks
and its future dynamical evolution. The summary and conclusions resulting
from this research are outlined in the last section.

\begin{center}
\begin{table*}
\caption{List of all observatories involved in observing
comet C/1890~F1 Brooks. The last observatory (Shattuck Observatory
operated by Dartmouth College in Hanover, New Hampshire) was not present
in the standard MPC~list of observatories at the time when we performed our calculations. Therefore, it had our new code
marked by negative number: $-58$.  Now, this observatory is included in the list and its code is 307. 
In our basic set of data we omitted: seven transit observations from Washington, four unpublished observations
from Li\`{e}ge, and additional nine from Vienna as they were published after the monumental
Str\"{o}mgren's \citeyearpar{stromgren:1896} paper, see text for more
details.} \label{tab:observatories}

\setlength{\tabcolsep}{1.0pt} 
\begin{tabular}{lclc}
\hline 
Observatory code \& place  & Number of  & Observatory code \& place  & Number of \tabularnewline
 & measurements  &  & measurements \tabularnewline
\hline 
000 Greenwich  & 62  & 004 Toulouse  & 17 \tabularnewline
007 Paris  & 5  & 008 Algiers-Bouzar\'{e}ah  & 13 \tabularnewline
014 Marseilles  & 25  & 020 Nice  & 18 \tabularnewline
035 Copenhagen  & 21  & 045 Vienna  & \qquad{}78\,$+$\,9 \tabularnewline
084 Pulkovo  & 54  & 085 Kiev  & 58 \tabularnewline
089 Nikolaev  & 9  & 135 Kasan  & 33 \tabularnewline
513 Lyons  & 3  & 516 Hamburg  & 72 \tabularnewline
522 Strasbourg  & 8  & 526 Kiel  & 13 \tabularnewline
528 G\"{o}ttingen  & 36  & 529 Christiania  & 27 \tabularnewline
531 Collegio Romano, Rome  & 12  & 532 Munich  & 26 \tabularnewline
533 Padua  & 15  & 537 Urania Obs., Berlin  & 13 \tabularnewline
539 Kremsmünster  & 38  & 548 Berlin  & 6 \tabularnewline
582 Orwell Park Obs., Ipswich  & 15  & 601 Dresden, Engelhardt Obs.  & 13 \tabularnewline
623 Li\`{e}ge  & 4  & 662 Lick Obs., Mount Hamilton  & 13 \tabularnewline
753 Washburn Obs., Madison  & 6  & 767 Ann Arbor  & 16 \tabularnewline
780 Leander Mc Cormick Obs., Charlottesville  & 8  & 787 U.S. Naval Obs., Washington  & \qquad{}22\,$+$\,7 \tabularnewline
794 Vassar College Obs., Poughkeepsie  & 9  & 802 Harvard Obs., Cambridge  & 12 \tabularnewline
999 Bordeaux-Floirac  & 99  & -58 Hanover, Shattuck Obs., New Hampshire  & 13 \tabularnewline
\hline 
\end{tabular}
\end{table*}

\par\end{center}

\begin{center}
\begin{table*}
\caption{Global description of all available astrometric
observations of C/1890~F1 Brooks. Our basic set of collected data
consists of 888 positional measurements that cover the same time-interval
as given in the fourth column (see text for an explanation).}\label{tab:Obs-mat} 

\setlength{\tabcolsep}{4.0pt} 
\begin{tabular}{lcccrcc}
\hline 
Comet  & q$_{{\rm obs}}$  & T  & Observational arc  & No  & Data  & Heliocentric \tabularnewline
name  &  &  & dates  & of  & arc span  & distance span \tabularnewline
 & {[}au{]}  & {[}yyyy\,mm\,dd.ddddd{]}  & {[}yyyy\,mm\,dd -- yyyy\,mm\,dd{]}  & obs  & {[}yr{]}  & {[}au{]} \tabularnewline
\hline 
C/1890~F1 Brooks  & 1.908  & 1890\,06\,02.03838  & 1890\,03\,22 -- 1892\,02\,05  & 908  & 1.9  & 2.11 -- 6.56 \tabularnewline
\hline 
\end{tabular}
\end{table*}

\par\end{center}

\section{Observations of C/1890~F1 Brooks and Str\"{o}mgren's orbital determination}

\label{section:data}

Comet C/1890~F1 (called comet 1890a or 1890~II at that time) was discovered
at Smith Observatory, Geneva, N.Y., USA by William R.~Brooks
on 1890 March~19, at 16 hours standard, 75th meridian time (March 20.38~UT, see \cite{brooks:1890}) at a heliocentric
distance of 2.1\,au and 2.7\,au from the Earth. The comet was extensively
observed during next several months. In the first days of June
1890 the comet reached its perihelion at a distance of 1.91\,au from
the Sun (on June~2) and the day after -- the closest distance of
1.57\,au from the Earth.

The Cometography by Gary W.~\citet[vol.2, pages 648-652]{kronk2:2004}
was very helpful in collecting rich literature concerning comet C/1890~F1~Brooks.
It appeared that all these original papers are nowadays available
in an electronic form what highly accelerated our data processing. Thanks to
the NASA's Astrophysics Data System Bibliographic Services (ADS) 
\footnote{http://adswww.harvard.edu/
}, we were able to gather full original versions of almost all sources
listed by Kronk, except for the papers published in \emph{Comptes Rendus}.
The latter are available at the French \emph{Gallica} digital library
\footnote{http://gallica.bnf.fr/
}.

Additionally, an extensive study of C/1890~F1 Brooks published by
Elis~\citet{stromgren:1896} was extremely helpful in our investigation.
Unfortunately, this book is not available from the ADS, however it can be found
at \emph{Hathi Trust Digital Library} 
\footnote{https://www.hathitrust.org/
}. Str\"{o}mgren collected all available 
\footnote{After his investigation nine more measurements taken in Vienna were
published and these are also included in our orbit determination using
Tycho-2 star catalogue.%
} positions (899 in number) spanning the entire period of measurements.
Then, in contemporary catalogues he identified almost all reference
stars used by observers several years earlier, and on this basis he
recalculated almost all comet's positions. The final osculating orbit
obtained by him is based on 854 observations reduced again, carefully
weighted and then grouped in sixteen normal places spread over the
period from 1890~March~20 to 1892~February~05. Next, he added
approximate perturbations by Earth, Mars, Jupiter, and Saturn. This
solution is presented in many sources as the most credible
osculating orbit for this comet. For example, an osculating orbit
given in the last edition of the Catalogue of Cometary Orbits \citet[hereafter MWC08]{marsden-cat:2008}
is cited as an orbit derived by Str\"{o}mgren, however the epoch of orbit
is taken close to perihelion passage (1890 June~2), whereas Str\"{o}mgren
adopted the osculation epoch close to the beginning of data sequence
(1890 March~17). It means that osculating orbit derived originally
by Str\"{o}mgren was dynamically evolved a two and a half months forward
for the MWC08 purposes and finally transferred to the J2000 reference
frame.

In Table~\ref{tab:observatories},  the list of all observatories
involved in observing the comet C/1890~F1 according to Str\"{o}mgren's
paper is presented (36 observatories, page 31 therein). We independently collected
888 original cometary positions given by observers in their original
papers and this set of data is used here as 'basic set of original
cometary positions' (hereafter 'basic set of data') , which were next
recalculated using the PPM star catalogue.

\noindent One can see that almost 50\,per cent of all data were gathered
in six of presented observatories: Bordeaux~(99 measurements), Vienna~(87),
Hamburg~(72), Greenwich~(62), Kiev~(58) and Pulkovo~(54).

\begin{figure}
\includegraphics[width=8.8cm]{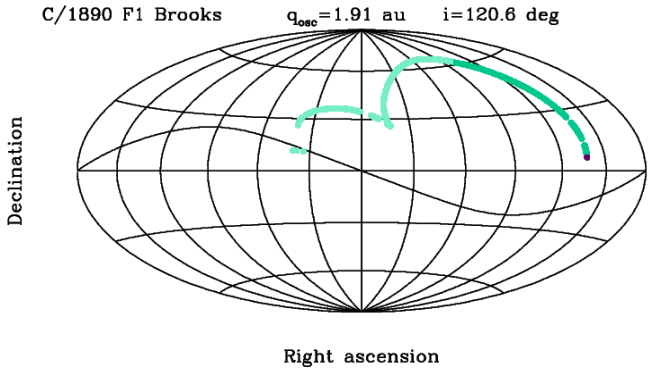}
\caption{The overall view on cometary track of
C/1890~F1 Brooks filled by collected astrometric observations in
geocentric equatorial coordinate system given in Aitoff projection.
Declination is plotted along the ordinate, right ascension is plotted
along the abscissa (increasing from zero to 360 degree from the left
to right) and the centre of projection is at 0\degr declination and
180\degr right ascension. The lines of right ascension and declination
are shown at 30\degr intervals and the wavy line shows the projection
of the ecliptic onto the celestial sphere. Each positional observation
before perihelion passage is shown as a dark turquoise point, except
the first observation that is shown by a dark magenta point whereas
the data taken after perihelion passage are marked by light turquoise
points.}\label{fig:geocentric_sky} 
\end{figure}

All the collected data are plotted in Fig.~\ref{fig:geocentric_sky},
where discovery position is given by magenta point, pre-perihelion
data are shown as dark-turquoise points, and light-turquoise points
follow the post-perihelion trajectory on the sky.

\begin{figure}
\includegraphics[width=8.8cm]{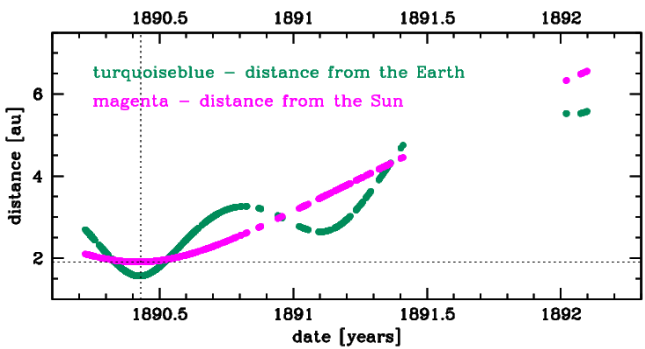}
\caption{The time distribution of positional observations
of C/1890~F1 Brooks with corresponding heliocentric (magenta curve)
and geocentric (turquoise curve) distance at which they were made.
Horizontal dotted line shows the perihelion distance and vertical
dotted line the moment of perihelion passage.}\label{fig:distances} 
\end{figure}

Global characteristics of the collected observational material are
given in Table~\ref{tab:Obs-mat}. The asymmetry in observed heliocentric
distances around perihelion is easily visible in Fig.~\ref{fig:distances}
(magenta track), where one can see that C/1890~F1 was discovered
at a distance of about 2.1\,au from the Sun and was followed far after
the perihelion passage (1.9\,au) to about 6.7\,au from the Sun.
In the entire dataset of C/1890~F1, there are only two photographic
measurements, taken by Charles Tr\'{e}pied on 1890~May~22 at Algiers
Observatory. It is also worth noting here that St\'{e}phane~Javelle (Nice
Observatory) was the only observer who followed the comet after the
seven-month gap in the data due to comet's conjunction with the Sun. 

\section{Two methods of data processing}

\label{section:star_recalculation}

In our orbit recalculation presented here, we primarily relied on original
data published by observers. We also partially based our analysis on the Str\"{o}mgren's~\citeyearpar{stromgren:1896}
paper where the whole set of data is published. However, we only used 
(\begin{tiny}\Pisymbol{astrosym}{34}\end{tiny}\textendash \,$\star$)-type of Bordeaux observations given there, and, optionally, corrections to all 99~cometary positions in declination taken in Bordeaux since both types of data were not published anywhere else. The author received them directly from Georges Rayet, the director of Bordeaux
observatory. Therefore, to be as close to Str\"{o}mgren's sample of observations
as possible, our 'basic set of data' consists of 888 positional observations,
all with given (\begin{tiny}\Pisymbol{astrosym}{34}\end{tiny}\textendash \,$\star$)-type of measurements. 
Our full sample of data includes twenty more observations (from Washington, Vienna
and Li\`{e}ge Observatories, see Tables~\ref{tab:observatories} and~\ref{tab:Obs-mat}).

The only intervention into the original data are our manual corrections
of typing mistakes which sometimes appear in published tabular data,
or explicit observer's errors manifesting as a kind of outliers
(e.g. distant from preliminary orbit by full number of minutes
in right ascension or/and declination). Some of them were identified
by Str\"{o}mgren, often by a direct contact with observers. Thus, we took
almost all of them into account and we also identified several more.
The complete list of the adopted corrections can be found at our \texttt{WikiComet} page
\footnote{\texttt{http://apollo.astro.amu.edu.pl/WCP}%
} in the supplementary material to this paper.

Next, we used direct measurements of comet positions relative to reference
stars, that is the (\begin{tiny}\Pisymbol{astrosym}{34}\end{tiny}\textendash \,$\star$)-type of data, in order to recalculate
new positions of the comet using more modern star catalogues. For
this purpose, we used two different approaches based on PPM and Tycho-2
catalogues, respectively.

\subsection{ Method I of comet position recalculations}

\label{section:PPM_procedure}

\label{stars_in_PPM}

\begin{center}
\begin{table*}
\caption{The process of preliminary determination of the osculating orbit: 
comparison between original data and data recalculated
using PPM star catalogue. For Bordeaux set of data (subset B, both
version) selection in steps 1a and 1b was limited to the rejection of
three largest outliers. In the remaining cases a sharp Bessel criterion
was applied for data selection during the process of orbit determination; 
the resulting cut-off level (all measurements that give larger residuals are not taken into account) 
is displayed in columns {[}3{]}/{[}6{]} and {[}9{]}.} \label{tab:PPM_reduction}

\setlength{\tabcolsep}{2.0pt} 
\begin{tabular}{lcrccrccrcc}
\hline 
Data  & Number of  & \multicolumn{3}{c}{B e f o r e ~~~P P M} & \multicolumn{6}{c}{A f t e r ~~~P P M}\tabularnewline
description  & measurements  & \multicolumn{3}{c}{{S t e p I} } & \multicolumn{3}{c}{{S t e p I I}} & \multicolumn{3}{c}{{S t e p I I I }}\tabularnewline
 &  & \multicolumn{3}{c}{~~ } & \multicolumn{3}{c}{using the same limit for outliers} & \multicolumn{3}{c}{repeated selection procedure}\tabularnewline
 &  & Level of  & Number of  & rms  & Level of  & Number of  & rms  & Level of  & Number of  & rms \tabularnewline
 &  & cut-off  & residuals  &  & cut-off  & residuals  &  & cut-off  & residuals  & \tabularnewline
$[1]$  & $[2]$  & $[3]$  & $[4]$  & $[5]$  & $[6]$  & $[7]$  & $[8]$  & $[9]$  & $[10]$  & $[11]$\tabularnewline
\hline 
subset A  & 789  & 17\farcs 0  & 1380  & 5\farcs 08  & 17\farcs 0  & 1421  & 4\farcs 71  & 14\farcs 7  & 1404  & 4\farcs 35 \tabularnewline
subset B, version O  & 99  & 25\farcs 0  & 197  & 6\farcs 95  & 25\farcs 0  & 198  & 6\farcs 11  & 15\farcs 1  & 193  & 5\farcs 41 \tabularnewline
subset B, version C  & 99  & 25\farcs 0  & 197  & 5\farcs 51  & 25\farcs 0  & 198  & 4\farcs 46  & 8\farcs 9  & 193  & 3\farcs 20 \tabularnewline
\hline 
\end{tabular}
\end{table*}

\par\end{center}

The first method used here is based on the reverse process to the one performed
by observers. The star position is calculated from the (\begin{tiny}\Pisymbol{astrosym}{34}\end{tiny}\textendash \,$\star$)-measurements listed by an observer and from a given cometary position and next corrected using the
PPM star catalogue \citep{gabryszewski:1997}. Sometimes in original papers only 
(\begin{tiny}\Pisymbol{astrosym}{34}\end{tiny}\textendash \,$\star$)-measurements 
are given (in right ascension and/or in declination). When comet's position is not given in one
or both of the coordinates, we temporarily calculate the values of
the missing coordinates which fit the cometary orbit, and then we
are looking for a matching star. If successful, we obtain a position
of the comet in the previously unknown one or two coordinates. This
method does not require any identification of individual reference star
and its mean position, because if we can find (in the PPM star catalogue in this realisation)
a star within a radius of few tens of arc seconds from the original
position adopted by an observer, then we are sure in practice, that this is
the correct star. More details about this procedure applied for recalculating
comet positions are described by \citet{gabryszewski:1997} and
next by \citet{krol-sit-et-al:2014} who presents the successful application of this approach
to 38~Oort spike comets discovered in the first half of the twentieth
century.

When a certain fraction of the reference stars is successfully found for
a given sample of observations, this procedure always leads to an
improvement of the orbit determination by (i) reduction of the root
mean square (rms) residual and/or by (ii) enlargement of the number of
applicable observations for orbit determination. Thus, it also leads to smaller uncertainties of
orbital elements. Therefore, we often use this method as an effective
tool for extensive studies based on large samples of comets. For this method we
decided to take a sample of 888~observations with
known (\begin{tiny}\Pisymbol{astrosym}{34}\end{tiny}\textendash \,$\star$)-measurements, including the set of 99~measurements
from Bordeaux. There are 38 observations where only
(\begin{tiny}\Pisymbol{astrosym}{34}\end{tiny}\textendash \,$\star$)-type of data are given in original papers, and
a few dozens more for which the difference in (\begin{tiny}\Pisymbol{astrosym}{34}\end{tiny}\textendash \,$\star$),
and as a consequence the cometary position, is given in only one coordinate
(in right ascension or in declination). Since Str\"{o}mgren recalculated comet's
positions for almost all measurements using more modern catalogues
than were previously available to the observers, it will be interesting
to compare here his osculating orbit of C/1890~F1 with our orbital
results based on automatic search for reference stars in the PPM 
catalogue, and next with the results obtained with a different re-reduction algorithm and Tycho-2 catalogue (Section~\ref{section:stars_in_Tycho-2}).

As we mentioned before, Str\"{o}mgren published the corrections to declinations
for Bordeaux measurements, which he received directly from Rayet  who observed C/1890~F1 five years earlier. 
Since the literal application of all these corrections does not seem to be obvious 
(see Section~\ref{section:stars_in_Tycho-2} for an in-depth discussion on these corrections), 
we divided the data into two subsets: one containing only Bordeaux measurements (subset 'B') 
and the other with all remaining observations 
from the basic set of 888 measurements (subset 'A'). Next, we applied
the PPM procedure to both, independently; and additionally we performed
this step for two variants of subset 'B' data. In version 'O',
we dealt with the original measurements, whereas in version 'C' the
corrections to $\Delta\delta$ published in pages 32--33 of Str\"{o}mgren's
paper were applied. The PPM~procedure gave the same results
for 'O' and 'C', what was expected. Namely, using the PPM~procedure,
corrected positions for the same 51 reference stars of 99~original measurements were found.
In the case of subset~A  containing 789 observations, the reference
star positions for 402~measurements were recalculated. Next, 
for each of three subsets of data (each containing about 50~per
cent of recalculated cometary positions), the same sharp Bessel
selection criterion was independently applied during the process of osculating orbit
determinations (for more details about our methods of data selection
see \citet{krolikowska-sit-soltan:2009}, and references therein).
The results are summarized in Table~\ref{tab:PPM_reduction} and
the main conclusions from these steps of data processing are as follows: 

\begin{itemize}
\item Automatic search procedure for reference stars in PPM \citep{gabryszewski:1997}
allowed us to find new star positions for about 50 per cent of measurements
in the case of C/1890~F1, while Str\"{o}mgren found manually about 92 per cent
of stars in catalogues available to him in 1896. 
The difference in the effectiveness of star search results not only from the manual approach
applied by Str\"{o}mgren, but also from the fact that the method discussed now, ignores, for purposes of comparison, 
the data on the stars used by the observers and Str\"{o}mgren. 
These original stellar data are used in our second method (see section~\ref{section:stars_in_Tycho-2}).
\item This recalculation of star positions significantly improved the rms
of the whole data sets -- compare columns {[}5{]} and {[}11{]} of
Table~~\ref{tab:PPM_reduction}. The most spectacular decrease of
rms was obtained for Bordeaux data when the Rayet corrections were
applied to $\Delta\delta$ measurements and, as a consequence, to
original comet positions in $\delta$ for these measurements. 
\item Rayet corrections to positions in declination appeared to significantly
improve the rms at each step of orbit determination (compare columns
{[}5{]}, {[}8{]} and {[}11{]} in the variant~C and variant~O) 
\end{itemize}
\noindent Since Rayet corrections to declinations significantly reduced
scattering of data points around the orbit determined from Bordeaux
measurements (these data are spread over a relatively long arc of orbit corresponding to the time interval  
of about 1.1\,yr),  we decided to construct two types of data recalculated
using the PPM~star catalogue: 
\begin{itemize}
\item DATA\_Ia where Rayet corrections were ignored, 
\item DATA\_Ib where Rayet corrections were applied, 
\end{itemize}
\noindent to show how the orbit changed under the influence of these 
$\Delta\delta$-corrections suggested by Rayet for Bordeaux measurements,
which represent about 11~per cent of the whole set of data and only in declination. 

At the end of this section it is interesting to trace the relative
weights of those observatories, where the largest number of data were
taken. We decided to weigh the measurements according to the place where they
were made and whether they are successfully converted using the PPM
catalogue or not. Therefore, Table~\ref{tab:6_obs_weights} (columns {[}1{]}--{[}5{]})
gives relative weights for subsets of original measurements,
that is for those where stars were not found in PPM catalogue (column {[}4{]}),
and for subsets where stars were found and cometary positions were
recalculated (column {[}5{]}). We found that for all these subsets
of measurements the recalculated parts are less dispersed around the
cometary orbit than those consisting of original positions. This fact
is manifested in higher weights in column~{[}5{]} than in column~{[}4{]}
for each observatory given in Table~\ref{tab:6_obs_weights}. A spectacular
improvement of comet's position residuals is found for Pulkovo
Observatory (Russia). In fact, astrometric measurements taken by Franz
Renz using the 38-centimetre (15-inch) refracting telescope are the
best in the whole data set for C/1890~F1.

Summarizing, Tables~\ref{tab:PPM_reduction}--\ref{tab:6_obs_weights}
show that the use of automatic search for star position in modern star catalogue
(here: PPM catalogue); starting only with comet's positions and \mbox{(\begin{tiny}\Pisymbol{astrosym}{34}\end{tiny}
\textendash\,$\star$)}-measurements in hand
significantly improves positional observations of comets discovered
long ago.

\begin{center}
\begin{table*}
\caption{Relative weights for six observatories
with the largest number of measurements for three versions of data
handling: DATA\_Ib (columns {[}4{]} \& {[}5{]}), DATA\_IIa (column
{[}6{]}) and DATA\_IIb (column {[}7{]}).}\label{tab:6_obs_weights} 

\setlength{\tabcolsep}{5.0pt} 
\begin{tabular}{ccccccc}
\hline 
Observatory  & Number of  & Per cent  & \multicolumn{4}{c}{R e l a t i v e ~~~w e i g h t s}\tabularnewline
location  & all  & of obs.  & \multicolumn{2}{c}{D a t a\_Ib~~-- ~weighted ~solution} & \multicolumn{2}{c}{D a t a\_II~-- ~weighted ~solution}\tabularnewline
 & obs.  & recalculated & original part   & part of obs.  & \multicolumn{2}{c}{recalculated using Tycho-2 catalogue}\tabularnewline
 &       & using PPM    & of obs. (star   & recalculated  & corr. for Bordeaux & corr. for Bordeaux    \tabularnewline
 &       & catalogue    & in PPM not found) & using PPM   & omitted            & included              \tabularnewline
$[1]$  & $[2]$  & $[3]$ & $[4]$  & $[5]$  & $[6]$  & $[7]$ \tabularnewline
\hline 
Bordeaux  & 99  & 50.5  & 0.738  & 0.795  & 0.529  & 1.232 \tabularnewline
Vienna  & 78  & 62.8  & 0.515  & 0.604  & 0.630  & 0.623 \tabularnewline
Hamburg  & 72  & 54.2  & 0.869  & 1.031  & 1.121  & 0.821 \tabularnewline
Greenwich  & 62  & 48.4  & 0.422  & 0.515  & 0.372  & 0.318 \tabularnewline
Kiev  & 58  & 46.6  & 0.609  & 0.619  & 0.683  & 0.561 \tabularnewline
Pulkovo  & 54  & 66.7  & 1.458  & 3.406  & 3.874  & 3.602 \tabularnewline
\hline 
\end{tabular}
\end{table*}

\end{center}

\begin{figure*}
\caption{O-C diagrams for comet C/1890~F1 Brooks
with DATA\_IIb-version of observations. Two upper panels present the
time distribution of positional observations with corresponding residuals
based on unweighted data (orbital solution~A1) and weighted data (orbital
solution~A5), respectively, where residuals in right ascension are
shown as green dots and in declination as red triangles. The lowest
panel shows relative weights for solution~A5. Note
that the horizontal axes show the time elapsed from the beginning
of the 1890 AD.}\label{fig:O-C_diagram}
\includegraphics[width=17cm]{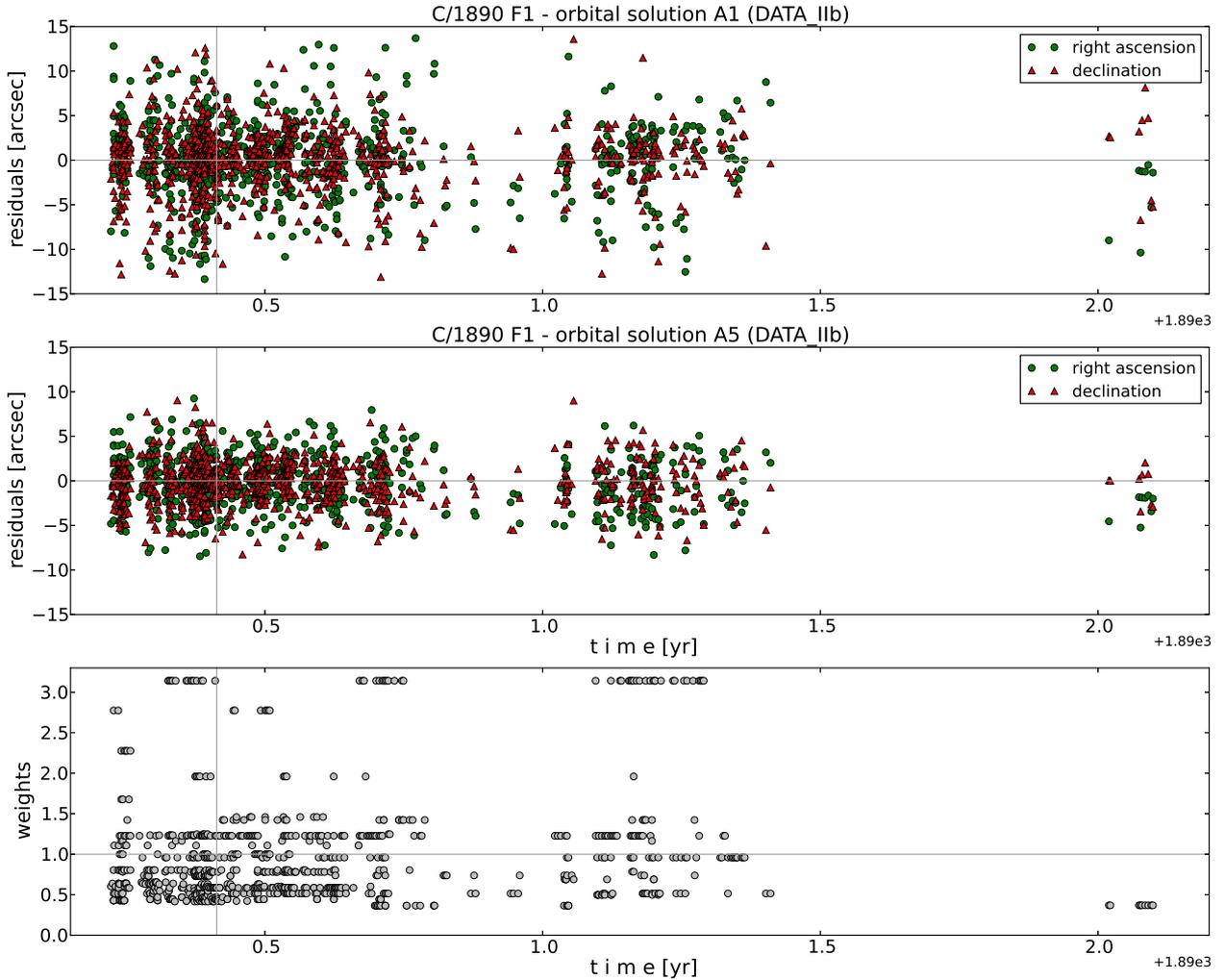} 
\end{figure*}

\subsection{Method II. In-depth method of comet position recalculation using Tycho-2 catalogue}

\label{section:stars_in_Tycho-2}

The second, partly automatic and therefore more time consuming method of positional
measurements handling was initially inspired by the monumental work
of Elis Str\"{o}mgren \citeyearpar{stromgren:1896}, which has been already mentioned
in previous sections. The main idea is generally the same: use
only the (\begin{tiny}\Pisymbol{astrosym}{34}\end{tiny}\textendash \,$\star$)-measurement, take mean coordinates of the reference star used by
the observer and calculate a comet position using this star position
taken from the most precise source. Str\"{o}mgren collected 899 observations
of C/1890~F1 Brooks and completed a list of 486 different reference
stars used by observers. It seems worth to mention that for 17 stars
Str\"{o}mgren himself calculated proper motions which were unavailable at that time.
On the other hand, he rejected 40~stars as having no reliable catalogue
positions available to him and, as a result, he discarded 45 observations
based on these stars. In his paper he noted that additional five observations
were published after his work was completed, and in fact we found
nine such additional observations, all from Vienna observatory.
We found all the original papers containing C/1890~F1 observations, except
for four observations made by dr~De~Baal at Li\`{e}ge Observatory, which
Str\"{o}mgren obtained in a private communication and which probably were never
published separately. In the case of these four observations we used
data given in Str\"{o}mgren's paper.

Since our aim here was to recalculate cometary positions using 
(\begin{tiny}\Pisymbol{astrosym}{34}\end{tiny}\textendash \,$\star$)-type 
of measurements and contemporary stellar data, we
excluded  seven meridian circle observations made
at Washington \citep{superintendent:1890} from this procedure. We only reconstructed the 
moments of these observations taken from this
original communication. Additionally, a problem with Bordeaux observations also appeared:
among 99 measurements only in five cases the published data contained differences 
(\begin{tiny}\Pisymbol{astrosym}{34}\end{tiny}\textendash \,$\star$). We decided to use the differences presented by \citet{stromgren:1896},
which he reconstructed from original publications or obtained (along
with the above mentioned corrections) in the course of personal
communication with Georges Rayet, the Bordeaux Observatory director
at that time.

Our first step was to supply a list of mean stellar positions copied
from Str\"{o}mgren's list of reference stars, as a search target for the
SIMBAD database 
\footnote{http://simbad.u-strasbg.fr/simbad/
}. Str\"{o}mgren's stellar data were so good that 88~per cent of stars
were automatically found within the radius of 10\,arcsec. Another 11~per cent (50 of 446 stars)
was found in the distance between 10 and 30\,arcsec, and only a few
stars in distances larger than 30\,arcsec from the positions given
by Str\"{o}mgren. However, there were numerous ambiguous identifications,
and of course not all of the indubitable identifications were correct, so
a lot of manual search and corrections was necessary. Next, we
tried to identify those 40~stars for which Str\"{o}mgren presented no coordinates,
and all of them were found in SIMBAD database by using information given
by observers in their original papers. We also found seven additional
stars used for nine Vienna observations unavailable for Str\"{o}mgren
because these measurements were published after his paper was.
In all, we obtained a list of 491 SIMBAD identifications of the reference
stars used by C/1890~F1 observers (surprisingly, there are two stars from Str\"{o}mgren's
list that were not used by any observer). We were interested in the most precise
astrometric parameters of these stars. About 40~per cent of them
could be found in the Hipparcos Catalogue \citep{hipparcos-cat:1997}
but due to the importance of proper motions and the necessity to use
internally consistent data, we decided to use Tycho-2 catalogue as
a source of positions and proper motions. We found Tycho-2 identifications
for all of 491 reference stars, and only for five of these stars there are
no solution given in this catalogue (these stars are marked with the
'X' flag - \textquotedbl{}no mean position, no proper motion\textquotedbl{}),
so we decided to use positions and proper motions from the UCAC4 catalogue
\citep{zacharias-UCAC4:2012} for them. This addition
to Tycho stars is so small that in the remaining text we call the
stellar data source simply the Tycho-2 catalogue. 

In order to avoid unnecessary calculations for individual observations,
we have introduced the following algorithm. 
\begin{itemize}
\item First, we corrected a reference star position 
for its proper motion from the standard epoch of J2000 to the moment of observation. 
These corrections were to be considered for the relatively large time-interval
of 110~years and we used also parallaxes and radial velocities from SIMBAD
database whenever they were available. 
\item Second, we applied an appropriate precession matrix to express star
coordinates in a mean reference frame of the epoch of observation. 
\item Next, we added the measured (\begin{tiny}\Pisymbol{astrosym}{34}\end{tiny}\textendash \,$\star$) differences. 
\item At the end, we reversed the precession calculation from the second
step above. 
\end{itemize}
As a result a list of 901 astrometric, topocentric positions
of a comet was obtained. Moreover, they were expressed in the J2000 ICRS frame and ready for orbit
determination. Such a procedure allows us to omit, for example, corrections
for nutation, aberration etc. At the end we added seven meridian observations
from Washington (with the epochs of observations reconstructed from
local sidereal times), making the total number of observations
in DATA\_II set equal to 908. Among them 84 observations consisted of
only one coordinate (right ascension or declination) what made a
number of 1732~residuals potentially suitable for orbit determination.

\begin{center}
\begin{table*}
\caption{Osculating orbits of C/1890 F1 Brooks
determined by \citet{stromgren:1896} and by the present investigation.
Our orbital elements of osculating heliocentric orbits are given for
three alternative sets of data and two methods of further data processing.
The successive columns present: $[1]$ -- Epoch of osculation, type of solution (A1: unweighted
data, A5: weighted data), and quality of orbit (1a or 1b) estimated
using modified method of quality assessment \citep{kroli-dyb:2013},  $[2]$ -- perihelion time {[}TT{]},
$[3]$ -- perihelion distance, $[4]$ -- eccentricity, $[5]$ -- argument
of perihelion (in degrees), equinox J2000.0, $[6]$ -- longitude of
the ascending node (in degrees), equinox J2000.0, $[7]$ -- inclination
(in degrees), equinox J2000.0, $[8]$ -- reciprocal semi-major axis
in units of $10^{-6}$\,AU$^{-1}$, and $[9]$ -- rms and number of residuals used for orbit determination. Osculating orbits determined
for weighting data are indicated by light grey shading.}\label{tab:orbits_osculating}

\setlength{\tabcolsep}{3.0pt} 
\begin{tabular}{crrrrrrrr}
\hline 
Epoch $[$yyyymmdd$]$  & T$_{{\rm obs}}$  & $q_{{\rm obs}}$  & $e_{{\rm obs}}$  & $\omega_{{\rm obs}}$  & $\Omega_{{\rm obs}}$  & $i_{{\rm obs}}$  & $1/a_{{\rm obs}}$  & rms, \tabularnewline
Type of solution,  & $[$yyyymmdd.dddddd$]$  & $[$AU$]$  &  & $[$\degr$]$  & $[$\degr$]$  & $[$\degr$]$  & $[10^{-6}$AU$^{-1}]$  & Number of \tabularnewline
Quality of orbit  &  &  &  &  &  &  &  & residuals \tabularnewline
$[1]$  & $[2]$  & $[3]$  & $[4]$  & $[5]$  & $[6]$  & $[7]$  & $[8]$  & $[9]$ \tabularnewline
\hline 
 &  &  &  &  &  &  &  & \tabularnewline
\multicolumn{9}{c}{\textbf{O s c u l a t i n g ~~o r b i t ~~d e t e r m i n e d ~~b y ~~E l i s ~~S t r \"{o} m g r e n}}\tabularnewline
\multicolumn{9}{c}{\textbf{(original Rayet corrections for Bordeaux measurements of $\Delta\delta$ are included)}}\tabularnewline
 &  &  &  &  &  &  &  & \tabularnewline
18900317  & 18900602.033026  & 1.90758325  & 1.00041030  & 68.934397  & 320.345283  & 120.556094  & \multicolumn{2}{c}{\citet{stromgren:1896} }\tabularnewline
class: 1A  & $\pm$0.000447  & $\pm$0.00000307  & $\pm$0.00001300  & $\pm$0.000231  & $\pm$0.000096  & $\pm$0.000064  & \multicolumn{2}{c}{equator \& equinox: 1890}\tabularnewline
18900602  & 18900602.037500  & 1.90758200  & 1.00026600  & 68.927300  & 321.877900  & 120.569000  & \multicolumn{2}{c}{MWC08}\tabularnewline
 &  &  &  &  &  &  &  & \tabularnewline
\hline 
 &  &  &  &  &  &  &  & \tabularnewline
\multicolumn{9}{c}{\textbf{ Present investigation on the basis of 888 measurements and PPM star catalogue}}\tabularnewline
 &  &  &  &  &  &  &  & \tabularnewline
\multicolumn{9}{c}{\textbf{T y p e ~~~~o f ~~~~d a t a : ~~~~~DATA\_Ia}}\tabularnewline
 &  &  &  &  &  &  &  & \tabularnewline
\rowcolor{LightGray} 18900602  & 18900602.037418  & 1.90757866  & 1.00030730  & 68.927555  & 321.877730  & 120.568830  & $-$161.09  & 2\farcs 94 \tabularnewline
\rowcolor{LightGray} \textcolor{red}{A5}, class: 1a  & $\pm$0.000294  & $\pm$0.00000217  & $\pm$0.00000763  & $\pm$0.000145  & $\pm$0.000043  & $\pm$0.000047  & $\pm$ 4.00  & 1591 \tabularnewline
 &  &  &  &  &  &  &  & \tabularnewline
\multicolumn{9}{c}{\textbf{T y p e ~~~~o f ~~~~d a t a : ~~~~~DATA\_Ib }}\tabularnewline
\multicolumn{9}{c}{\textbf{(original Rayet corrections for Bordeaux measurements of $\Delta\delta$ are included)}}\tabularnewline
 &  &  &  &  &  &  &  & \tabularnewline
18900602  & 18900602.037782  & 1.90757411  & 1.00027856  & 68.927521  & 321.877534  & 120.568981  & $-$146.03  & 4\farcs 30 \tabularnewline
\textcolor{cyan}{A1}, class: 1b  & $\pm$0.000471  & $\pm$0.00000329  & $\pm$0.00001251  & $\pm$0.000241  & $\pm$0.000086  & $\pm$0.000070  & $\pm$ 6.56  & 1598 \tabularnewline
\rowcolor{LightGray} 18900602  & 18900602.037418  & 1.90757887  & 1.00030640  & 68.927548  & 321.877721  & 120.568824  & \textbf{$-$160.62}  & 2\farcs 89 \tabularnewline
\rowcolor{LightGray} \textcolor{blue}{A5}, class: 1a  & $\pm$0.000291  & $\pm$0.00000215  & $\pm$0.00000758  & $\pm$0.000143  & $\pm$0.000043  & $\pm$0.000047  & \textbf{$\pm$ 3.97}  & 1592 \tabularnewline
\hline 
 &  &  &  &  &  &  &  & \tabularnewline
\multicolumn{9}{c}{\textbf{ Present investigation on the basis of 908 measurements and Tycho-2 star catalogue}}\tabularnewline
 &  &  &  &  &  &  &  & \tabularnewline
\multicolumn{9}{c}{\textbf{T y p e ~~~~o f ~~~~d a t a : ~~~~~DATA\_IIa}}\tabularnewline
 &  &  &  &  &  &  &  & \tabularnewline
\rowcolor{LightGray} 18900602  & 18900602.037518  & 1.90757958  & 1.00032677  & 68.927671  & 321.877701  & 120.568916  & $-$171.30  & 3\farcs 18 \tabularnewline
\rowcolor{LightGray} \textcolor{red}{A5}, class: 1a  & $\pm$0.000274  & $\pm$0.00000200  & $\pm$0.00000699  & $\pm$0.000133  & $\pm$0.000039  & $\pm$0.000043  & $\pm$ 3.66  & 1681 \tabularnewline
 &  &  &  &  &  &  &  & \tabularnewline
\multicolumn{9}{c}{\textbf{T y p e ~~~~o f ~~~~d a t a : ~~~~~DATA\_IIb }}\tabularnewline
\multicolumn{9}{c}{\textbf{(linear corrections for Bordeaux measurements of $\Delta\delta$
are included)}}\tabularnewline
 &  &  &  &  &  &  &  & \tabularnewline
18900602  & 18900602.038077  & 1.90757261  & 1.00029412  & 68.927720  & 321.877503  & 120.569119  & $-$154.18  & 3\farcs 95 \tabularnewline
\textcolor{cyan}{A1}, class: 1b  & $\pm$0.000421  & $\pm$0.00000292  & $\pm$0.00001121  & $\pm$0.000217  & $\pm$0.000078  & $\pm$0.000061  & $\pm$ 5.88  & 1659 \tabularnewline
\rowcolor{LightGray} 18900602  & 18900602.037682  & 1.90758152  & 1.00033162  & 68.927751  & 321.877680  & 120.568896  & $-$173.84  & 2\farcs 61 \tabularnewline
\rowcolor{LightGray} \textcolor{blue}{A5}, class: 1a  & $\pm$0.000236  & $\pm$0.00000173  & $\pm$0.00000658  & $\pm$0.000116  & $\pm$0.000037  & $\pm$0.000037  & $\pm$ 3.45  & 1644 \tabularnewline
 &  &  &  &  &  &  &  & \tabularnewline
\hline 
\end{tabular}
\end{table*}

\par\end{center}

Preliminary calculation of the O-C~residuals with respect to the
Str\"{o}mgren's orbit revealed numerous large residuals. Deep inspection
of each case allowed us to eliminate most of the suspected gross
errors. In this process, we confirmed most of the corrections proposed
by Str\"{o}mgren (see notes at pages 70--71 of his paper), sometimes finding
different explanations. We also introduced a certain number of new corrections, eliminating
some incorrect reference star identifications or various typos found
in original papers.

Special attention was given to the corrections of north polar
distance (NPD) for all Bordeaux observations, published by Str\"{o}mgren
after he had received them from Rayet through personal correspondence. Str\"{o}mgren
quoted the Rayet explanation for these corrections: ``Les observations
de la Com\`{e}te 1890 II ont \'{e}t\'{e}, \`{a} l'origine, r\'{e}duites avec une valeur
inexacte du tour de la vis de d\'{e}clinaison de l'\'{e}qvatorial''.  All these corrections were carefully checked
and it appeared that they can be divided into two groups: most of them are linear functions of the observed NPD differences (as expected) but about 10~per cent seems to be somehow modified. To discriminate between these two groups we performed 
least square fitting that allowed us to obtain a simple formula
for \textbf{linear} corrections: 
\begin{equation}\label{equ_linear}
NPD_{\textrm{cor}}=0.016981\times\Delta NPD+0.010516,
\end{equation}
where $\Delta NPD$ is a published difference in NPD measured by an
observer and $NPD_{\textrm{cor}}$ is the calculated correction and
both are expressed in arc seconds.
After a detailed inspection we decided to apply all corrections that are linear with respect to the observed NPD as fully legitimate and  eight individual, additional corrections proposed by   Str\"{o}mgren or found by us as probable typing errors in original papers. The list of all corrections applied by us (not only for the Bordeaux observations) can be found in an auxiliary material to this paper, see the note at the end of this paper.

Similarly as in section~\ref{section:PPM_procedure}, we constructed here two versions of data (with and without corrections to Bordeaux (\begin{tiny}\Pisymbol{astrosym}{34}\end{tiny}\textendash \,$\star$)-measurements) recalculated 
with the use of Tycho-2 catalogue to study their overall influence on the resulting comet orbit: 
\begin{itemize}
\item DATA\_IIa where corrections to $\Delta\delta$ of Bordeaux measurements
were ignored, 
\item DATA\_IIb where corrections to $\Delta\delta$ of Bordeaux measurements
were applied. 
\end{itemize}
However, in the DATA\_IIb set we decided to apply corrections for
Bordeaux declinations calculated from the linear model (described
above) instead of those listed by Str\"{o}mgren and based on Rayet's estimations.

Columns 6 and 7 of Table~\ref{tab:6_obs_weights} show that the Bordeaux
subset of data is about 2.5 times less scattered along the orbit when
these corrections for $\Delta\delta$ are applied. As a result Bordeaux
subset of data in the DATA\_IIb (column~7) has greater weights than
subsets of data from Vienna, Hamburg, Greenwich and Kiev, while without
these corrections the situation is reversed.

\section{New osculating orbit determinations}

\label{section:our_osculating_orbits}

In the previous section, we constructed two versions of data for each of
two methods of data recalculations. We are convinced that the most reliable  osculating orbit is based
on data recalculated with the use of the Tycho-2 star catalogue, including
the corrected Bordeaux $\Delta\delta$-measurements according to linear formula given by Eq.~\ref{equ_linear}, and involving the deep data processing to eliminate as many sources of errors as possible (a solution based on DATA$\_$IIb).

However, to put the problem of orbit determination into a wider perspective,
we constructed here a grid of osculating orbit starting from $2\times2$
sets described in Section~\ref{section:star_recalculation}. Then,
for each of them we performed two types of final data treatment during
the process of orbit determination: 
\begin{itemize}
\item selection procedure based on Bessel criterion (hereafter called A1
type of solution), 
\item selection and weighting procedure (hereafter called A5 type of solution);
for our methods of data weighting see \citet{krolikowska-sit-soltan:2009}
or \citet{kroli-dyb:2010}. 
\end{itemize}
It should be emphasized that the process of data selection (and weighting)
is individually performed for each data set and each type of model
of motion (ballistic or non-gravitational), separately. In
other words, it is always carried out simultaneously during the iterative
process of osculating orbit determination. In this paper we present only purely
gravitational solutions (ballistic) because non-gravitational effects are very
poorly determinable (pre-perihelion branch of orbit covered by data
is distributed only over a two-months period and perihelion distance reaches 1.9\,au, see also
a very narrow range of the observed heliocentric distances before
perihelion in Fig.~\ref{fig:distances}). Thus, we decided to not include
such uncertain solutions in the presented here analysis.

The grid of final osculating orbits is presented in Table~\ref{tab:orbits_osculating}
and the O-C-diagram for DATA\_IIb-version of observations is shown
in Fig.~\ref{fig:O-C_diagram}. In Table~\ref{tab:orbits_osculating} we omitted only
two unweighted solutions for data sets constructed without the Bordeaux corrections. 
Nevertheless, these solutions are shown in Fig.~\ref{fig:O-C_diagram} with magenta symbols. 
All osculating orbits given in Table~\ref{tab:orbits_osculating}
are of the 1A quality class using \citet{mar-sek-eve:1978} quality
class assessment. However according to a modified method that we have recently
introduced \citep{kroli-dyb:2013}, we noticed differences in
quality classes: solutions based on unweighted data are of 1b~class,
whereas solutions based on weighted observations are of 1a~class
(see column {[}1{]} of the table). We consider the orbit based
on the weighted data which was previously recalculated with the use of the Tycho-2
star catalogue with linear corrections for Bordeaux measurements of
$\Delta\delta$ (solution A5, DATA\,IIb) the best osculating orbit given here. 
Therefore, this osculating
orbit was next used in the analysis of dynamical evolution of C/1890~F1
presented in section~\ref{section:past_next_orbits}. Additionally,
the (O-C)-distribution is Gaussian only in this case.

Fig.~\ref{fig:O-C_diagram} shows the gap in observations toward
the end of data set that stretches from 1891 May~29 to 1892 Jan~7 (seven
months). After that gap, the comet was observed only by S.~Javelle
from Nice Observatory who took nine positional measurements in the period from
January~7 to February~5, as was mentioned in section~\ref{section:data}.
Those days the comet was more than 6\,au from the Sun, and more than
5.5\,au from the Earth. He noted that the comet 'was extremely weak,
very badly defined, and one minute wide at most' (citation after Cometography, Kronk 2013).
In Fig.~\ref{fig:O-C_diagram} one can see that these
observations are well-distributed around the orbital solution in declination. However
all measurements in right ascension give negative residuals regardless
whether the orbit was determined using unweighted (upper panel) or
weighted data (middle panel, note small weights of this set of observations).
On the other hand, these measurements are very important, because
they extend the period of observations by more than eight months and
thereby determine the good quality of the orbit. Apart from this trend in
right ascension at the end of data, no other systematic trends in
the remaining residuals are seen in the O-C~diagram based on weighted
data. As for the modern standards, large scatter of residuals
around the orbit based on unweighted data (upper panel of Fig.~\ref{fig:O-C_diagram})
draws our attention. It was, however, not unusual at that time.

\begin{figure}
\caption{Projection of the 6D~space of 5\,001\, clones of C/1890~F1 onto six
chosen planes of osculating orbital elements for solution A5 obtained
using our best version of observations (DATA\_IIb. i.e. recalculated
using Tycho-2 star catalogue and applying linear corrections for Bordeaux measurements
of $\Delta\delta$). Vertical axes given in the right-hand panels
are exactly the same as the vertical axes in the left-hand panels.
Each grey point represents a single virtual orbit, while the large
blue points with solid error bars represent the nominal weighted orbital
solution for DATA\_IIb given in Table~\ref{tab:orbits_osculating}.
The analogous solution derived using weighted observations in the
DATA\_Ib-version are given also by blue dots inside the dotted error
bars. The remaining solutions given here are shown using red dots
(solutions A5: DATA\_IIa, DATA\_Ia), magenta squares (A1: DATA\_IIa,
DATA\_Ia), cyan squares (A1: DATA\_IIb, DATA\_Ib), and green triangles
which represent the solution obtained by Str\"{o}mgren. The zero point of each axis is centred
on the nominal values of the respective pair of osculating orbital
elements denoted by the subscript '0' for the best solution (A5, DATA\_IIb; see also Table~\ref{tab:orbits_osculating})
and error bars show 1$\sigma$ errors.}\label{fig:osculating_swarms_DATAIIb} 
\includegraphics[width=8.8cm]{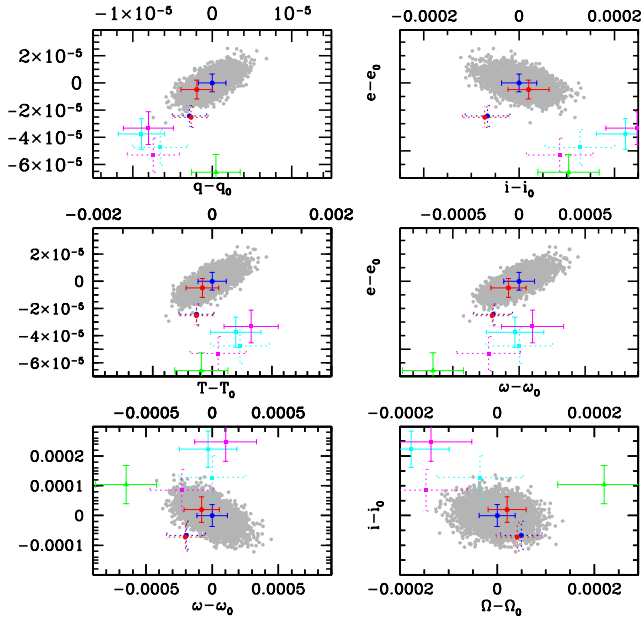} 
\end{figure}

Fig.~\ref{fig:osculating_swarms_DATAIIb} compares our best osculating
orbit (solution A5 based on DATA\_IIb, blue point with solid blue
error bars in the figure) with the rest of grid of osculating orbits
derived here, and with Str\"{o}mgren solution given in MWC\,08, that
is recalculated for the same epoch of 1890 June~2 (and standard equator and equinox:
2000.0) as all of our solutions discussed here (see also Table~\ref{tab:orbits_osculating}).
This figure presents six two-dimensional projections which indicate
to what extend all these solutions are compatible. From the comparison of
all the solutions shown in Fig.~\ref{fig:osculating_swarms_DATAIIb}
and given in Table~\ref{tab:orbits_osculating} we can draw following
conclusions. 

\begin{itemize}
\item All osculating orbits, including the solution obtained by Str\"{o}mgren,
are consistent within 3-4$\sigma$-combined error, whereas all our
weighted solutions are consistent within 1-2$\sigma$-combined error
(blue and red points). 

\item The differences in orbital elements between respective pairs of solutions
that differ only in ignoring or including the $\Delta\delta$-corrections
to Bordeaux measurements, are always deeply below 1$\sigma$-combined
error for each of the orbital elements (compare respective pairs of blue
and red dots, and magenta and cyan dots). The pairs of respective orbits
derived using weighted data based on PPM catalogue (blue and red dots
with dotted error bars) are closer to each other than pairs of weighted
data based on Tycho-2 catalogue (blue and red dots with solid error
bars). In order to understand this behaviour we recommend to
analyse the weights given in Table~\ref{tab:6_obs_weights} for Bordeaux
observations and for different versions of data. 

\item Orbital element uncertainties obtained by Str\"{o}mgren are very similar
to those derived by us using the unweighted data (A1-type of solutions,
cyan and magenta error bars). 

\item Osculating orbits based on weighted data are characterized by significantly
smaller orbital element uncertainties than those given by Str\"{o}mgren.
He obtained his solution by using the $\Delta\delta$-correction for
Bordeaux measurements, and somehow weighting the data when compressed
into 16~normal points. Thus, we conclude that both methods used
herein and based on two different modern catalogues give significantly better quality of orbits. 

\item Using the Tycho-2 catalogue to recalculate all measurements of C/1890~F1,
and applying all necessary corrections, we also derived the best quality
of osculating orbit (solution A5, DATA\_IIb shown in Fig.~\ref{fig:osculating_swarms_DATAIIb}
as a blue dots with solid error bars). 

\item The method based on automatic search for stars in PPM~star catalogue proposed by
\citet{gabryszewski:1997} allowed to recalculate 50~per cent of
measurements. This also significantly improves the orbit quality and
provides a solution (blue point with dotted error bars) within 2$\sigma$-combined
error from our best one. 
\end{itemize}

\section{Original and future barycentric orbits}

\label{section:original_future_orbits}

\begin{center}
\begin{table}
\caption{Original and future barycentric inverse
semimajor axes for grid of orbital solutions of C/1890~F1. Values
determined using solutions based on weighting data are indicated by light grey shading.}
\label{tab:original_future_1/a}
\setlength{\tabcolsep}{1.0pt} 
\begin{tabular}{cccc}
\hline 
Solution  & $1/a_{{\rm ori}}$  & $1/a_{{\rm fut}}$  & $1/a_{{\rm fut}}-1/a_{{\rm ori}}$ \tabularnewline
type  & \multicolumn{3}{c}{{[}i n ~~~u n i t s ~~~o f ~~~au$^{-6}${]}}\tabularnewline
\hline 
 &  &  & \tabularnewline
\multicolumn{4}{c}{\textbf{T y p e ~~~~o f ~~~~d a t a : ~~~~~DATA\_Ia}}\tabularnewline
 &  &  & \tabularnewline
A1  & 79.35 $\pm$ 6.84  & 116.15 $\pm$ 6.84  & 36.796 $\pm$ 0.004 \tabularnewline
\rowcolor{LightGray} A5  & 67.05 $\pm$ 3.96  & 103.85 $\pm$ 3.96  & 36.798 $\pm$ 0.002 \tabularnewline
 &  &  & \tabularnewline
\multicolumn{4}{c}{\textbf{T y p e ~~~~o f ~~~~d a t a : ~~~~~DATA\_Ib}}\tabularnewline
 &  &  & \tabularnewline
A1  & 82.13 $\pm$ 6.63  & 118.93 $\pm$ 6.63  & 36.794 $\pm$ 0.004 \tabularnewline
\rowcolor{LightGray} A5  & 67.60 $\pm$ 3.95  & 104.40 $\pm$ 3.95  & 36.797 $\pm$ 0.002 \tabularnewline
 &  &  & \tabularnewline
\multicolumn{4}{c}{\textbf{T y p e ~~~~o f ~~~~d a t a : ~~~~~DATA\_IIa}}\tabularnewline
 &  &  & \tabularnewline
A1  & 71.77 $\pm$ 6.25  & 108.59 $\pm$ 6.24  & 36.802 $\pm$ 0.004 \tabularnewline
\rowcolor{LightGray} A5  & 56.91 $\pm$ 3.71  & ~~93.71 $\pm$ 3.71  & 36.803 $\pm$ 0.002 \tabularnewline
 &  &  & \tabularnewline
\multicolumn{4}{c}{\textbf{T y p e ~~~~o f ~~~~d a t a : ~~~~~DATA\_IIb}}\tabularnewline
 &  &  & \tabularnewline
A1  & 74.02 $\pm$ 5.85  & 110.82 $\pm$ 5.85  & 36.800 $\pm$ 0.003 \tabularnewline
\rowcolor{LightGray} A5  & 54.37 $\pm$ 3.42  & ~~91.18 $\pm$ 3.42  & 36.805 $\pm$ 0.002 \tabularnewline
\hline 
\end{tabular}
\end{table}

\end{center}

\begin{figure}
\includegraphics[width=8.8cm]{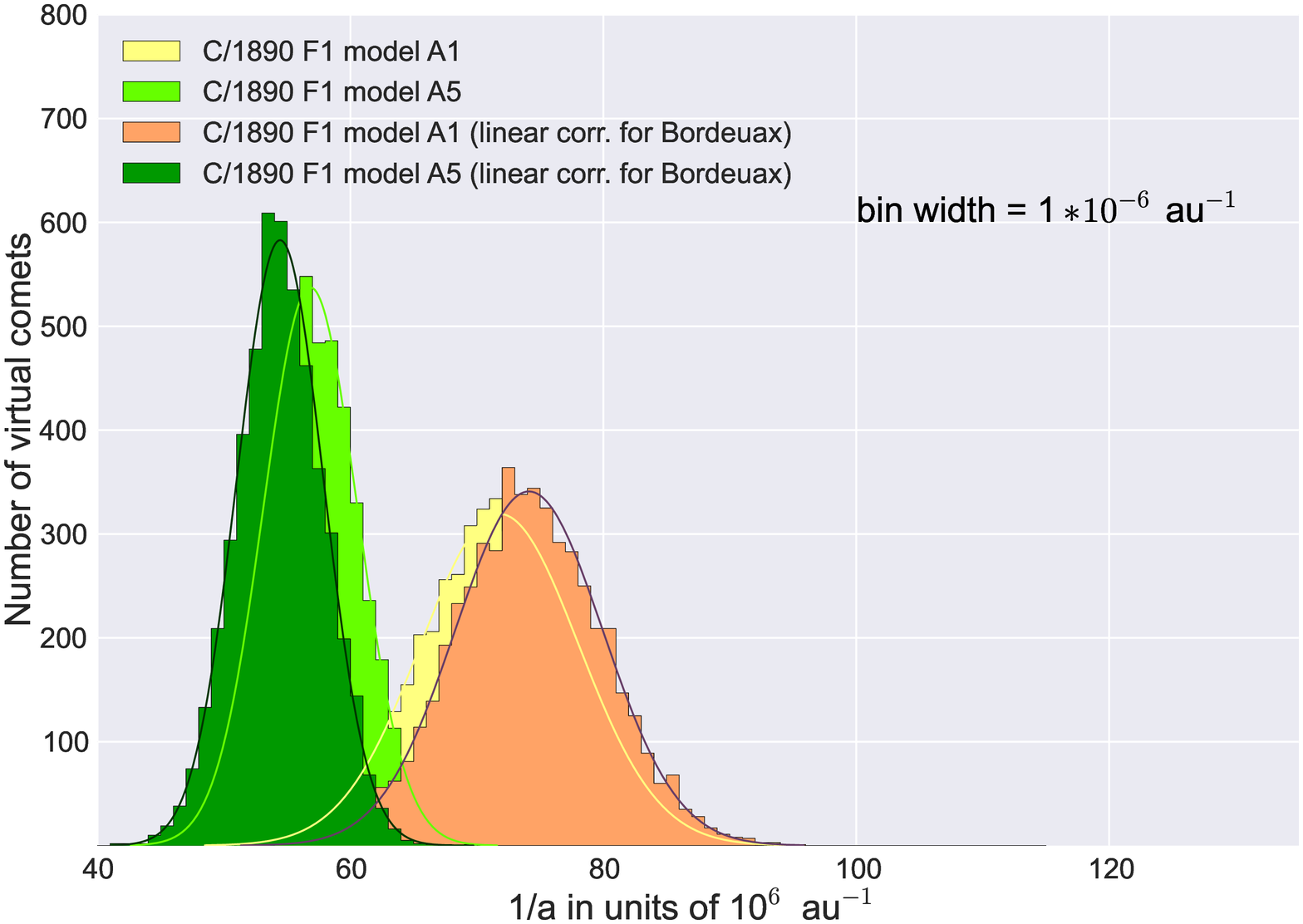}
\includegraphics[width=8.8cm]{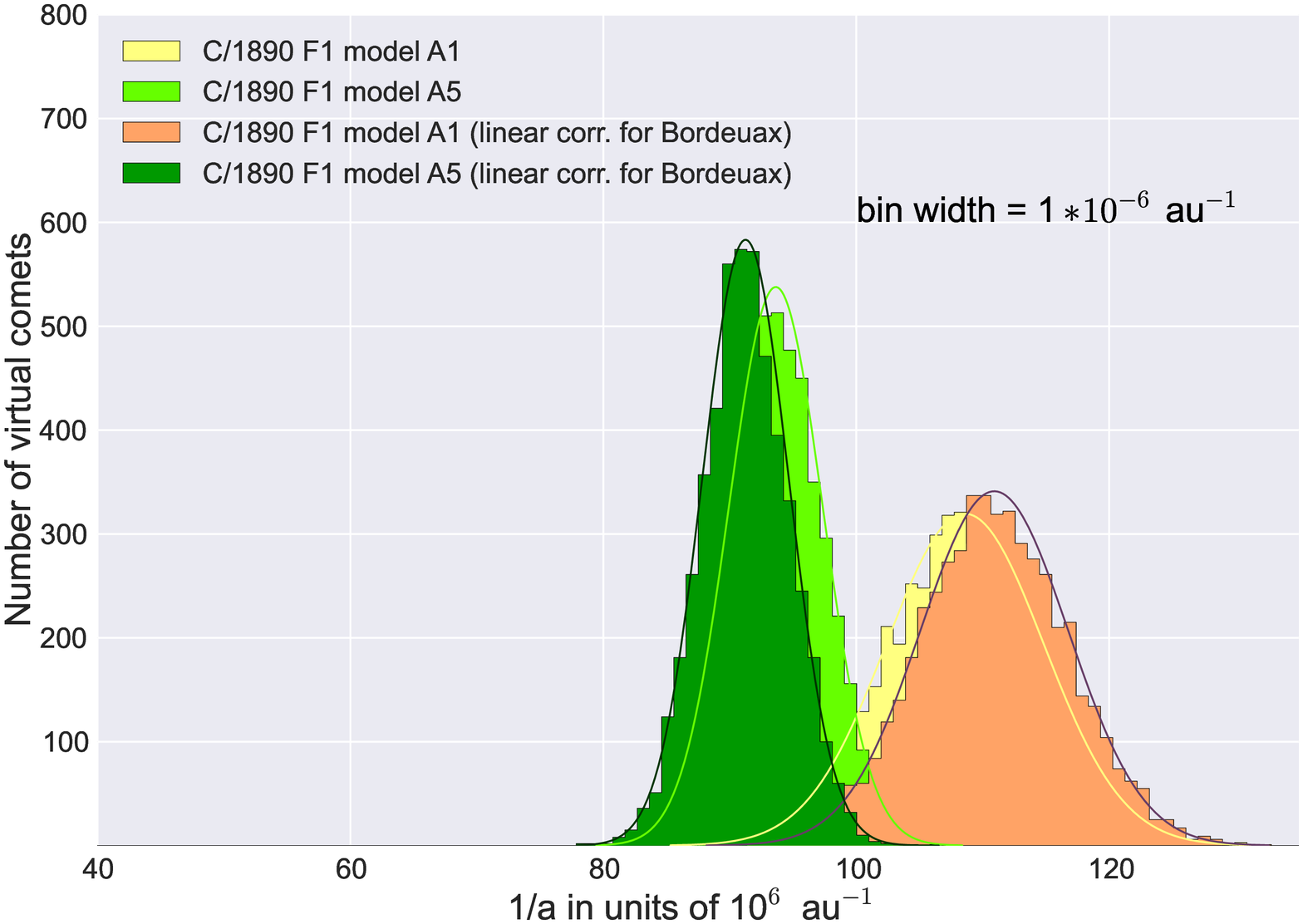}
\caption{Original (upper panel) and future (lower panel) $1/a$-distributions
for solutions based on data recalculated with the use of the Tycho-2-catalogue.
Each histogram represents the distribution of 5001 virtual comets (VCs), where the
Gaussian function (continuous line) gives perfect fit in each case.
The vertical axis shows counts in each bin within the sample of 5001
clones considered in each case; it means that counts of 500\,VCs
gives a probability of 0.1.}
\label{fig:original_axes_all_models} 
\end{figure}

\begin{figure}
\includegraphics[width=9.6cm]{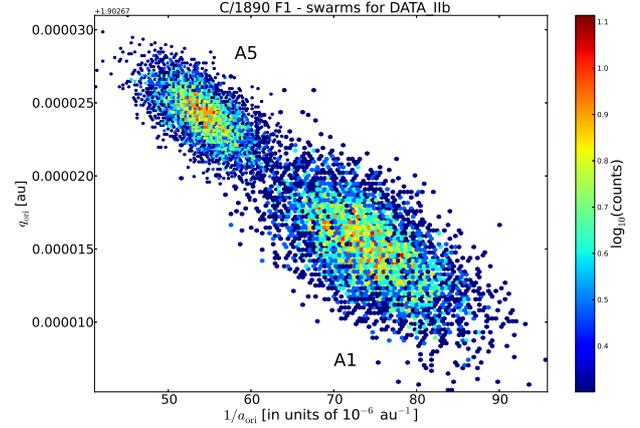}
\caption{Projection of two original swarms of 5001\,VCs (6D each) of C/1890~F1
Brooks onto the $1/a$--$q$ plane. These swarms represent solutions derived
from DATA\_IIb type of data, where the less dispersed swarm (upper left, and given by dark-green histogram in Fig.~\ref{fig:original_axes_all_models})
represents the A5-solution based on data which were weighted during
the procedure of orbit determination, whereas the more disperse swarm
shows A1-solution (given by sandy-brown histogram in the upper-left
panel of Fig.~\ref{fig:original_axes_all_models}). Density distribution
of A5-swarm is superimposed on the more dispersed A1-swarm. Density
map is given in logarithmic scale which is presented in the individual panel given on the right. 
Note that the value of the beginning of vertical
axis is given just above the upper horizontal border of the plot.}
\label{fig:original_swarms_DATAII} 
\end{figure}

To be able to reliably follow the evolution of cometary orbit we must
also know the orbital uncertainty 
\footnote{that results from the observational uncertainty of the osculating
orbit, see Table~\ref{tab:orbits_osculating}
} at each moment of our numerical calculations. Therefore for each
solution given in Table~\ref{tab:orbits_osculating} a swarm of
5001\,VCs, including the nominal orbit, was constructed according
to a Monte Carlo method given by \citet{sitarski:1998}, for more
details see also \citet{krolikowska-sit-soltan:2009}. Next, the dynamical
calculations of each swarm of VCs were performed backwards and forwards
in time until each VC reached 250\,au from the Sun, that is, a distance
where planetary perturbations are completely negligible.
This method allowed us to determine the uncertainties of original
and future orbital elements by fitting any orbital element distribution in the swarm
to Gaussian distribution at each moment of dynamical
evolution.

All the original and future barycentric $1/a$-values are given in Table~\ref{tab:original_future_1/a}
whereas the full original and future orbits are presented at \texttt{http://ssdp.cbk.waw.pl/LPCS}
and \texttt{http://apollo.astro.amu.edu.pl/WCP}. The distributions
of these original and future swarms of $1/a$ are shown in Fig.~\ref{fig:original_axes_all_models},
where the Gaussian fits to these distributions are also plotted. We
conclude that the best solution (DATA\_IIb, version A5) gives the values of 54.37$\pm$3.42 and 91.18$\pm$3.42,
in units of $10^{-6}$\,au$^{-1}$, for original and future $1/a$,
respectively. Both are the smallest values among all respective solutions
given in Table~\ref{tab:original_future_1/a}. \citet{stromgren:1914}
gives a value of 71.8 in the same units for original $1/a$, and this
value was cited by \citet{sinding:1948}, and next it was taken by Oort
for his famous analysis of $1/a_{\rm ori}$-distribution.

The mutual proximity of swarms A1 and A5 obtained using Tycho-2 (DATA\_IIb)
in two-dimensional projection is visualized in Fig.~\ref{fig:original_swarms_DATAII}.
Similar overlapping of the swarms we can find in every other projection.

It is worth noting that C/1890~F1 is very interesting from the planetary
perturbations point of view: it suffered very small perturbations
from planets during its passage through a planetary zone in the observed 1890-1892 apparition.
Regardless of the version of data used, these perturbations are at the level
of $\delta(1/a)=1/a_{{\rm fut}}-1/a_{{\rm ori}}\simeq36.8\times10^{-6}$\,au$^{-1}$
(last column of Table~\ref{tab:original_future_1/a}). As a result,
according to the best solution, the comet will be still inside the
Oort spike in the next apparition.

Table~\ref{tab:other_comets} gives other examples of comets that were subjected to weak planetary perturbations.
All were selected from the sample of about 160~Oort spike comets investigated by us so far. 
Here, we only present comets with small perihelion distance of $q_{{\rm obs}}<3.5$\,au, just
as we have in the case of comet C/1890~F1. However, we have also recognized
another 25~objects of that kind with large $q_{{\rm obs}}$ which constitute
a significant percentage of these large perihelion objects
(for more details see discussion in \citet{dyb-kroli:2011}). 
Otherwise, thirteen comets presented here (twelve from Table~\ref{tab:other_comets} plus C/1890~F1)
make up only about 15\,per cent of small perihelion Oort spike comets.
Among them five objects have orbits almost perpendicular to the ecliptic
plane and move on retrograde orbits, these are shown using light grey
shading in Table~\ref{tab:other_comets}. When we further reduce the sample
to comets that have passed to more bound orbit during the
passage through planetary zone we find only six such cases. Therefore, we can say that
because of suffering such a small planetary perturbations the dynamical
behaviour of C/1890~F1 is unusual among small perihelion comets.

\begin{center}
\begin{table*}
\caption{Original and future barycentric inverse
semimajor axes and their $\Delta1/a=1/a_{{\rm fut}}-1/a_{{\rm ori}}$ differences
for other Oort spike comets subjected to a weak planetary perturbations
during observed perihelion passage through the planetary zone ($|\Delta1/a|<100$
in units of $10^{-6}$au$^{-1}$). Comets with perihelion distance
closer than 3.5\,au to the Sun are only presented here. Original and future
$1/a$-values were taken from \citet{krolikowska:2014} and \citet{krol-sit-et-al:2014}.
Comets with inclination to ecliptic inside the range of
80\degr -- 100\degr (all on retrograde
orbits, see column {[}7{]}) are indicated by light grey shading.}\label{tab:other_comets}

\setlength{\tabcolsep}{6.0pt} 
\begin{tabular}{lccccccc}
\hline 
Comet  & $1/a_{{\rm ori}}$  & $1/a_{{\rm fut}}$  & $\Delta(1/a)$  & 1$\sigma$-error  & quality  & $q_{{\rm obs}}$  & $i_{{\rm obs}}$ \tabularnewline
 & \multicolumn{4}{c}{i n ~~~~~~~~u n i t s ~~~~~~~~o f ~~~~~~~~$10^{-6}$au$^{-1}$} & class  & {[}au{]}  & degrees \tabularnewline
$[1]$  & $[2]$  & $[3]$  & \multicolumn{2}{c}{$[4]$} & $[5]$  & $[6]$  & $[7]$ \tabularnewline
\hline 
C/1913 Y1 Delavan  & $52.57\pm4.23$  & $86.84\pm4.23$  & $+$34.27  & $<0.01$  & 1a  & 1.1  & 68.2 \tabularnewline
C/1919 Q2 Metcalf  & ~$34.7\pm67.3$  & $-26.9\pm67.3$  & $-$61.56  & ~~~~0.05  & 2a  & 1.1  & 46.4 \tabularnewline
C/1946 P1 Jones  & $50.86\pm5.08$  & $22.55\pm5.08$  & $-$28.32  & $<0.01$  & 1a  & 1.1  & 57.0 \tabularnewline
C/1946 U1 Bester  & $17.08\pm6.52$  & $43.88\pm6.52$  & $+$26.80  & $<0.01$  & 1b  & 2.4  & 108.2 \tabularnewline
C/1947 Y1 Mrkos  & $28.89\pm8.49$  & $54.31\pm8.50$  & $+$25.42  & $<0.01$  & 1b  & 1.5  & 77.5 \tabularnewline
\rowcolor{LightGray} C/1948 E1 Pajdu\v{s}\'{a}kov\'{a}-Mrkos  & $37.25\pm2.74$  & $35.21\pm2.74$  & ~~$-$2.04  & $<0.01$  & 1a  & 2.1  & 92.9 \tabularnewline
\rowcolor{LightGray} C/1952 W1 Mrkos  & $-0.1\pm85.8$  & $-41.4\pm106.9$  & $-$40.6~~  & ~43.3  & 2a  & 0.8  & 97.2 \tabularnewline
\rowcolor{LightGray} C/1989 Q1 Okazaki-Levy-Rudenko  & $42.9\pm22.2$ & $80.5\pm27.3$  & $+$37.6~~  & ~46.6  & 2a  & 0.6  & 90.1 \tabularnewline
\rowcolor{LightGray} C/1997 J2 Meunier-Dupouy  & $44.64\pm0.88$  & $14.72\pm0.91$  & $-$29.92  & ~~~~1.47  & 1a+  & 3.1  & 91.3 \tabularnewline
C/2006 HW$_{51}$ Siding Spring  & $47.31\pm3.37$  & $90.12\pm3.37$  & $+$42.81  & $<0.01$  & 1a  & 2.3  & 45.8 \tabularnewline
\rowcolor{LightGray} C/2006 S2 LINEAR  & $72.52\pm8.14$  & $-10.77\pm18.09$ & $-$83.3~~  & ~25.5  & 1b  & 3.2  & 99.0 \tabularnewline
C/2007 Q3 Siding Spring  & $39.13\pm0.49$  & $118.96\pm0.96$  & $+$79.83  & $<0.01$  & 1a+  & 2.3  & 65.7 \tabularnewline
\hline 
\end{tabular}
\end{table*}

\par\end{center}

\section{Previous and next perihelion passages}

\label{section:past_next_orbits}

With original and future barycentric orbits at hand, we can
study past and future motion of C/1890~F1 Brooks by following its
motion numerically. At distances greater than 250\,au we ignore all
planetary perturbations but we include Galactic and stellar perturbations
and integrate the comet motion to the previous and next perihelion
passage, that is about 2.5 Myr to the past and 1.1 Myr to the future.
Since previous and next perihelion passages of C/1890~F1 are deep
in the planetary zone and we cannot calculate planetary perturbations
at these epochs, we have to stop our calculations at these moments.
Of course for that reason one should treat our previous orbit as the
osculating one when the comet left the planetary zone after the previous
perihelion. Analogously one should treat the next orbit as the osculating one before
the comet will enter the planetary zone during its next apparition.

In this investigation we used exactly the same dynamical model as in Paper~5. It includes
Galactic disc and Galactic Centre tidal terms and perturbations from
90 stars or stellar systems known to pass closer than 3.5\,pc from
the Sun in the past, currently or in the future. Since we study C/1890~F1
motion in an interval smaller than 4\,Myrs these stellar data can
be considered complete in terms of massive and slow moving perturbers,
see Paper~5 for the detailed discussion and the description of the
stellar perturbers list. Stellar perturbers, together with a comet, are
numerically integrated as the N-body problem in the Solar system barycentric
frame with additional Galactic tidal potential. This calculation has
been repeated for the nominal orbit of this comet and for all 5000\,VCs
to propagate observational uncertainties.

\begin{figure}
\includegraphics[angle=270,width=0.9\columnwidth]{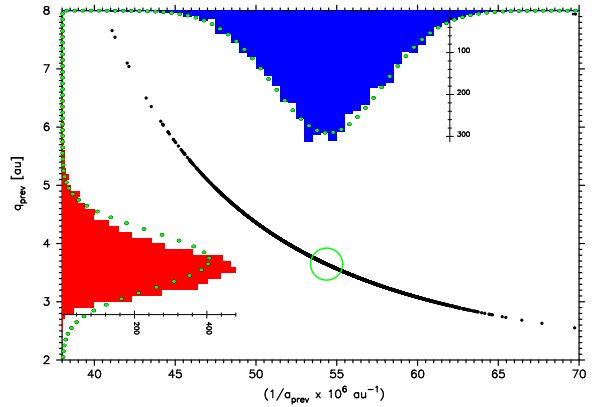} 
\caption{Joint and marginal distributions of
the osculating $1/a$ and $q$ of 5001 VCs derived from the nominal
orbit of  the preferred solution for C/1890~F1, 
stopped at their previous perihelion passage.
The centre of the green circle marks the nominal values. Small
green dots overprinted on the marginal distributions show best-fitting
Gaussian distributions. }\label{fig:Joint-and-marginal}
\end{figure}

Past and future motion of C/1890~F1 is quite regular, all previous
and next orbital clones move on elliptical orbits in each of the analysed
swarms. The most important parameters of these orbits are given in
Table \ref{tab:previous_and_next}. We present here the results of
two different calculations - with and without stellar perturbations
- in order to show how small their influence is. In contrast to almost perfectly
normal distributions of the inverse semimajor axes in all of the analysed
solutions, the perihelion and aphelion distances have non-Gaussian
distributions (see also Fig. \ref{fig:Joint-and-marginal}). Therefore,
in Table~\ref{tab:previous_and_next} we describe the distribution
of $q_{{\rm prev}}$ , $q_{{\rm next}}$, $Q_{\textrm{prev}}$ and
$Q_{\textrm{next}}$ by presenting their median values accompanied with 10th
and 90th deciles.

\begin{figure}
\includegraphics[angle=270,width=0.9\columnwidth]{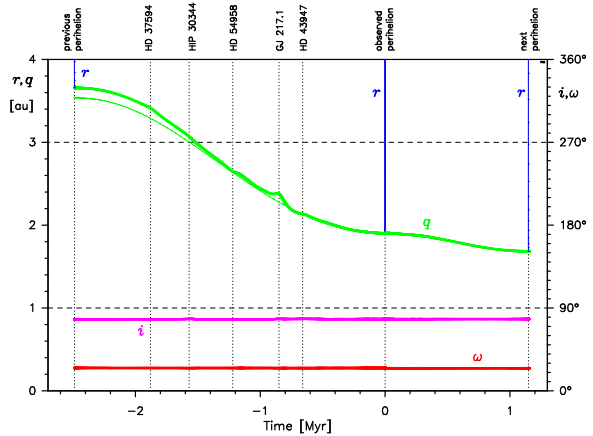}
\caption{Past and future evolution of C/1890~F1
nominal orbit of the preferred solution under the simultaneous Galactic and stellar perturbations.}
\label{fig:Nominal-C/1890F1-orbit}
\end{figure}

As it was highlighted in Section~\ref{section:original_future_orbits}
comet C/1890~F1 Brooks suffered only small planetary perturbations
during its observed perihelion passage. Looking in a wider perspective
we see that it was observed in a decreasing phase of its perihelion
distance evolution due to the Galactic tides. The perihelion distance
decreases from $q_{{\rm {prev}}}=3.64$~au through $q_{{\rm {obs}}}=1.91$~au
down to $q_{{\rm {next}}}=1.69$~au. Taking into account only small
semimajor axis shortening by planetary perturbations, C/1890~F1 is an example
of the Oort spike comet visiting deep interior of the Solar system
(and therefore being observable) in at least three consecutive perihelion
passages, still being a member of the spike. This behaviour is clearly
shown in Fig.~\ref{fig:Nominal-C/1890F1-orbit}.

\begin{center}
\begin{table*}
\caption{C/1890~F1 previous and next orbital parameters
derived from DATA\_IIb set of observation. The inverse semimajor
axis is presented as the Gaussian mean value and its 1$\sigma$ uncertainty;
$T_{\textrm{prev}}$ and $T_{\textrm{next}}$ are epochs of the previous
and next perihelion respectively. Perihelion and aphelion distances,
as well as their epochs are presented as 10th : 50th (the median) :
90th-deciles due to departure of their distributions from a normal
one. Solution name subscripted with a star means that all known stellar
and Galactic perturbations were included in the numerical integrations,
in contrast to the solutions without subscripts, where only Galactic perturbations
were included. }\label{tab:previous_and_next}

\setlength{\tabcolsep}{2.0pt}
\begin{tabular}{ccccccccc}
\hline 
Solution  & $T_{\textrm{prev}}$ & $1/a_{\textrm{prev}}$  & $q_{\textrm{prev}}$ & $Q_{\textrm{prev}}$ & $T_{\textrm{next}}$ & $1/a_{\textrm{next}}$ & $q_{\textrm{next}}$ & $Q_{\textrm{next}}$\tabularnewline
          & {[}Myr{]} & {[}10$^{-6}$au$^{-1}${]}  & {[}au{]}  & {[}10$^{3}$au{]}  & {[}Myr{]} & {[}10$^{-6}$au$^{-1}${]}  & {[}au{]}  & {[}10$^{3}$au{]}\tabularnewline
\hline 
A5$_{*}$ & $-$2.81:$-$2.48:$-$2.20 & $+54.44\pm3.42$  & 3.17:3.64:4.35  & 34.0:36.8:40.0  & 1.07:1.15:1.24 & $+91.24\pm3.42$  & 1.65:1.69: 1.72  & 20.9:21.9:23.0\tabularnewline
A5       & $-$2.81:$-$2.48:$-$2.20 & $+54.44\pm3.42$  & 3.09:3.53:4.20  & 34.0:36.8:40.0  & 1.07:1.15:1.24 & $+91.24\pm3.42$  & 1.65:1.69: 1.72  & 20.9:21.9:23.0\tabularnewline
A1$_{*}$ & $-$1.83:$-$1.56:$-$1.35 & $+74.17\pm5.85$  & 2.27:2.42:2.67  & 24.5:27.0:30.0  & 0.78:0.86:0.96 & $+110.91\pm5.85$ & 1.76:1.79: 1.81  & 16.9:18.0:19.3\tabularnewline
A1       & $-$1.83:$-$1.56:$-$1.35 & $+74.12\pm5.85$  & 2.25:2.40:2.64  & 24.5:27.0:30.0  & 0.78:0.86:0.96 & $+110.92\pm5.85$ & 1.76:1.79: 1.82  & 16.9:18.0:19.3\tabularnewline
\hline 
\end{tabular}
\end{table*}

\end{center}

Many interesting aspects of the past and future orbit evolution of
C/1890~F1 under stellar and Galactic perturbations can be found in
Fig.\ref{fig:Nominal-C/1890F1-orbit}. In this picture the horizontal
time axis extends from the previous perihelion passage through the
observed apparition up to the next perihelion passage. The left vertical
axis is expressed in au and describes the osculating perihelion distance
evolution (q, green line), as well as the heliocentric distance plot
($r$, thin blue lines). Due to the time scale of this picture, the heliocentric
distance plots take a form of vertical blue lines exactly at perihelion
passage moments. The right vertical axis is expressed in degrees and
describes changes in the osculating inclination ($i$, magenta line)
and in the argument of perihelion ($\omega$, red line). Both of these angular
elements are expressed in the Galactic frame. All thick lines depict
dynamical evolution under joint stellar and Galactic perturbations,
while the thin lines mark the evolution with the stellar perturbations
excluded. Horizontal dashed lines draw attention to the beginning
of the second and fourth quarter of $\omega$, whose values (90\degr
and 270\degr) are important from the point of view of the Galactic
perturbations (crossing these lines coincides with perihelion distance
minimum). The vertical dashed lines show the closest approaches of
a comet with the star or stellar system, which name is placed at the
top of the picture. It is worth mentioning that the timing of the stellar
perturbation is not necessarily strictly aligned with this closest
approach moment -- it strongly depends on the geometry, since the final
heliocentric orbit change is a net effect of a stellar gravitational
action on both a comet and the Sun. The action of GJ 217.1 is a good
example of a frequent self-cancelling stellar action: the perturbation
gained during the approaching phase is then cancelled during the receding phase.
During the presented 3.5\,Myr interval of cometary evolution,
the Galactic and stellar perturbations are notable only in the perihelion
distance, while being infinitesimally small in angular elements. We have not
identified any profound stellar perturbation in future motion of C/1890\,F1
since it will depart from the Sun for only about 20\,000\,au and it will
complete its future orbital revolution (1.1\,Myr) about 0.3\,Myr
before the predicted close flyby of the star Gliese~710. Other stars
are too small and/or too distant to noticeably change the future orbit of this comet.

\begin{figure}
\includegraphics[width=0.9\columnwidth]{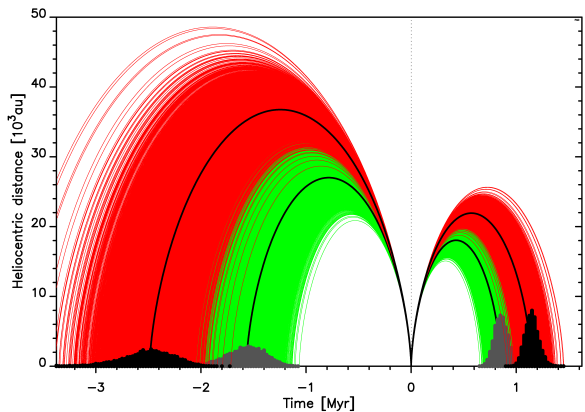} 
\caption{Past and future swarms of C/1890~F1 orbits
for two different solutions based on the DATA\_IIb set of observations:
red (dark grey) lines represent results for A5~solution (preferred, weighted
data) while green (light grey) lines correspond to the A1~solution
(without data weighting, see Tables \ref{tab:original_future_1/a}
and \ref{tab:previous_and_next} for more details). All known stellar
and Galactic perturbations were taken into account. Black lines depict
nominal orbits evolution. Distributions of previous and next perihelia
epochs are also shown.}\label{fig:fontanny}
\end{figure}

C/1890~F1 is definitely the dynamically old comet. Even the most
protruding clones in our swarm of VCs have the previous perihelion
distance well below 10\,au. The osculating perihelion distance and
the inverse semimajor axis distributions of C/1890~F1 for the moment
of the previous perihelion passage (nominally 2.5\,Myr ago) are presented
in Fig. \ref{fig:Joint-and-marginal}. It is worth mentioning that
the presented $1/a$-distribution is still almost perfectly Gaussian
while the $q$-distribution significantly departs from the normal one,
mainly as a result of Galactic perturbations. The central, black
joined distribution corresponds to the smaller cloud of 5001\,VCs presented
in Fig. \ref{fig:original_swarms_DATAII} after the 1.7--3.8\,Myr
of backward dynamical evolution under the simultaneous Galactic and
stellar action (different VCs have different orbital periods, a nominal
value equals 2.5\,Myr, the fastest VC was at the previous perihelion
1.7\,Myr ago while the slowest one almost 3.8\,Myr ago). 

The time-spread (along horizontal axis) and the space-spread (along
vertical axis) of our swarms of VCs in previous and next perihelion
are clearly visualized in Fig. \ref{fig:fontanny}. We presented here
changes in the barycentric distance of all 5001 VCs from two different
solutions of orbit determination, both based on the DATA\_IIb set
of observations. The red (darker) swarm depicts the distribution evolving
from the osculating orbit fitted to the weighted observations (A5-solution) while
the green (grey) one depicts the evolution of swarm constructed without
weighting (A1-solution). Nominal orbits for both of our solutions are marked as black
curves. Thus, this figure shows also how the data handling can change
the results during the evolution to the previous/next perihelion distance
(see also Table~\ref{tab:previous_and_next}). The dynamical evolution
of Str\"{o}mgren's orbit would be very similar to the evolution of our
A1 orbital solution. We conclude that our preferred A5-solution 
gives longer previous and next orbital periods of about 2.5 and 1.1\,Myr respectively. 
As a result, Galactic perturbations act more effective during the evolution, 
and for this solution we get a greater change of the perihelion distances 
between three consecutive perihelion passages analysed here.

\section{Summary and conclusions}

\label{section:conclusions}

A rich observational material of C/1890~F1 Brooks allows to test
whether it is productive to recalculate with the use of modern star catalogues 
the original positions of comets discovered long ago. During this study
we compared two distinct algorithms: 
\begin{itemize}

\item automatic search for reference stars in the PPM star catalogue according
to \citet{gabryszewski:1997}, which was frequently used by us when we only have (\begin{tiny}\Pisymbol{astrosym}{34}\end{tiny}\textendash \,$\star$)-measurements and data on the positions of comets,

\item automatic search for reference stars in the Tycho-2 star 
catalogue on the basis of mean coordinates of reference stars
used by all observers, and next, a detailed analysis of all the cases of 
large residuals in the  observed comet's positions, resulting from the star 
search ($\sim$30~per cent of all observations in the case of C/1890~F1).
\end{itemize}

The first method was successfully used several times before, and more recently
by \citet{krol-sit-et-al:2014}. This method of automatic search for stars in PPM
catalogue recalculates about 50~per cent of the existing data of C/1890~F1.
The second method was developed for the present investigation. In the SIMBAD database we
have successfully found all stars used by observers more
than hundred years ago to calculate the comet's position
in $\alpha$ and/or $\delta$ from (\begin{tiny}\Pisymbol{astrosym}{34}\end{tiny}\textendash \,$\star$)-measurements.
It means, that all positions of C/1890\,F1 were recalculated here
using these original (\begin{tiny}\Pisymbol{astrosym}{34}\end{tiny}\textendash \,$\star$)-measurements ($\Delta\alpha$
and/or $\Delta\delta$). We decided to use stellar positions and proper motions
from the Tycho-2 catalogue for that purpose.

\noindent It should be emphasized that thanks to the monumental publication
by Elis \citet{stromgren:1896}, our task has become much easier, though, it was
still time consuming. 
Searching for stars in the SIMBAD database would be
more difficult without the mentioned publication, particularly in the more complicated cases.
In addition, thanks to Str\"{o}mgren's deep analysis of data, it turned
out that one should take into account corrections to $\Delta\delta$
measurements from Bordeaux Observatory (we prefer linear version
of these corrections as described in section~\ref{section:stars_in_Tycho-2}). However, these corrections
change the orbital solution on the significantly lower level than
the use of modern stellar data. For example, a solution based on data
recalculated in about 50~per cent (PPM catalogue, DATA\_Ia/DATA\_Ib)
differs a lot from the respective solution based on fully recalculated
data (Tycho-2 catalogue DATA\_IIa/DATA\_IIb) than any two solutions
based on data with and without Bordeaux corrections. Moreover, the
Bordeaux corrections are also much less important than the method
of data treatment. This is clearly shown in Fig.~\ref{fig:original_axes_all_models},
where we can see that light-green histogram (weighted data, corrections
for Bordeaux omitted) is very similar to dark-green histogram (weighted
data, corrections for Bordeaux applied), however, it differs significantly
from the pair of yellow/orange histograms based on unweighted data.

Next, we have shown that both search algorithms for contemporary data for the reference 
stars resulted in a significant
reduction of the uncertainties of orbital elements in the case of
C/1890~F1 Brooks. The comparison of our solutions with that derived
by Str\"{o}mgren is presented in Table~\ref{tab:orbits_osculating} and
Fig.~\ref{fig:osculating_swarms_DATAIIb}. By analysing all these
models, we concluded that the orbital solution based on DATA\_IIb
(all positional observations were recalculated using Tycho-2 catalogue
and next weighted) gives the most reliable, and also the most accurate
osculating orbit.

Using this most preferred orbital solution (and its unweighted variant
for comparison, see Fig.~\ref{fig:fontanny}) we numerically followed
a motion of C/1890~F1 for one orbital period to the
past and to the future. Starting from full swarms of original and future
VCs orbits we obtained previous and next orbital elements of this
comet together with their uncertainties. All known stellar and Galactic
perturbations were fully taken into account but for comparison purposes
we have also performed the calculations in which stellar perturbations
were ignored. This comparison (see Table \ref{tab:previous_and_next}
for details) shows that none of known stellar perturbers significantly change
the past and future motion of C/1890~F1 during the three successive perihelion passages
analysed here. It should be stressed that due to relatively short past orbital period
of C/1890\,F1 Brooks (nominally 2.5\,Myr) it seems rather improbable
that we have missed here some unknown but significant (i.e. massive and/or
slow moving) and nearby stellar perturber.

Three consecutive perihelion passages of this comet (the time interval
of about 3.6\,Myr) clearly indicate that this comet deeply penetrates
the planetary region during each of them (below 5\,au from the Sun,
see Table~\ref{tab:previous_and_next}) and easily crosses the so-called
Jupiter-Saturn barrier. Therefore, we conclude that C/1890~F1 Brooks
is a dynamically old comet and additionally, as a result of negligible
planetary perturbations during the observed apparition, it will remain
a member of the Oort spike during the next apparition. Note that \citet{oort:1950}
treated this comet (by definition) as a dynamically new one and used
it in support for his cometary cloud hypothesis.

Currently, on the basis of analysis of 109~Oort spike comets, we estimate that
about 50~per cent of them are dynamically old. However, the percentage
of dynamically old comets grows up to almost 90 for comets with $1/a_{\rm ori}$
inside the range of 0.000040--0.000100\,au$^{-1}$ \citep{dyb-kroli:2015}.
Thus, the past dynamical evolution of comet Brooks having $1/a_{\rm ori}$
greater than $0.000050$\,au$^{-1}$ is not surprising. On the other
hand, for the reason of very weak planetary perturbations one can say that the
dynamical behaviour of C/1890~F1 is unusual among small perihelion
comets. 

In the near future, we plan to deal with all near-parabolic comets
discussed by \citet{oort:1950} as well as all Oort spike comets discovered
in the years 1901--1950 to answer the question about their source from
the previous perihelion perspective, and make our sample of near-parabolic
comets with known previous and next orbital elements more complete.
It can also happen that some comets observed on a more
tightly bound orbit ($1/a_{\rm obs}>0.000100$\,au$^{-1}$) will
contribute to a future $1/a$-distribution as Oort spike comets. Therefore
the extension of our research to objects with
$1/a$ similar to that shared by C/2013 V2 Borisov ($1/a_{\rm ori}=0.000770$\,au$^{-1}$and
$1/a_{\rm fut}=0.000089$\,au$^{-1}$, \citet{NK2952}) seems to
be important to put our investigation in the broader framework of
dynamical evolution of long-period comets.

Some auxiliary material to this paper is available at \texttt{http://ssdp.cbk.waw.pl/LPCS} and \texttt{http://apollo.astro.amu.edu.pl/WCP}.

\section{Acknowledgements}

We are very grateful to Ryszard Gabryszewski for providing software
and related assistance with the search for stars in PPM star catalogue.
We also thank the anonymous referee of this paper for valuable remarks.
\noindent This research has made use of NASA's Astrophysics Data System Bibliographic
Services and of the the SIMBAD database and VizieR catalogue access
tool, CDS, Strasbourg, France, and was partially supported from the project 2015/17/B/ST9/01790
founded by National Science Centre in Poland.







\bsp	
\label{lastpage}
\end{document}